\documentstyle[sprocl,epsf,rotate]{article}

\def\myfigure#1{\centerline{#1}}
\newcommand{\Foot}{\renewcommand{\thefootnote}{\fnsymbol{footnote}}
\footnote{\bf Lectures given at the
ICTP Summer School in Particle Physics (Trieste, 1999)}
\addtocounter{footnote}{-1}
\renewcommand{\thefootnote}{\alph{footnote}}}
\newcommand{\Preprint}{\vspace*{-2.5cm}  %hep-ph/0001XXX 
  \noindent \mbox{}\hfill IFIC/00-5  \\ \mbox{}\hfill FTUV/00-5 \\
  \vspace*{0.6cm} }
\renewcommand{\theequation}{\arabic{section}.\arabic{equation}}
\renewcommand{\arraystretch}{1.2}
\def\slashchar#1{\setbox0=\hbox{$#1$}\dimen0=\wd0%
\setbox1=\hbox{/}\dimen1=\wd1%
\ifdim\dimen0>\dimen1%                        
\rlap{\hbox to
\dimen0{\hfil/\hfil}}#1\else                                        
\rlap{\hbox to \dimen1{\hfil$#1$\hfil}}/\fi}
\def\Journal#1#2#3#4{{#1} {\bf #2}, #3 (#4)}
\def\add#1#2#3{{\bf #1}, #2 (#3)}

\def\NPB{{\em Nucl. Phys.} B}
\def\PLB{{\em Phys. Lett.}  B}
\def\PRL{\em Phys. Rev. Lett.}
\def\PRB{{\em Phys. Rev.} B}
\def\PRD{{\em Phys. Rev.} D}
\def\ZPC{{\em Z. Phys.} C}
\def\PR{\em Phys. Rev.}

\def\NPPS{\em Nucl. Phys. B (Proc. Suppl.)}
\def\PL{\em Phys. Lett.}

\def\JHEP{\em J. High Energy Phys.}
\def\etal{{\it et al.}}
\def\ie{{\it i.e.}}

\newcommand{\bel}[1]{\be\label{#1}}
\newcommand{\eqn}[1]{(\ref{#1})}
\newcommand{\be}{\begin{equation}}
\newcommand{\ee}{\end{equation}}
\newcommand{\ba}{\begin{array}{c}}
\newcommand{\bat}{\begin{array}{cc}}
\newcommand{\bath}{\begin{array}{ccc}}
\newcommand{\ea}{\end{array}}
\newcommand{\beqn}{\begin{eqnarray}}
\newcommand{\eeqn}{\end{eqnarray}}
\newcommand{\bi}{\begin{itemize}}
\newcommand{\ei}{\end{itemize}}

\newcommand{\dis}{\displaystyle}

\newcommand{\rms}{\rm\scriptsize}

\newcommand{\toU}{\stackrel{\mbox{\rms U(1)}}{\longrightarrow}}

\newcommand{\cL}{{\cal L}}

\newcommand{\cM}{{\cal M}}
\newcommand{\cO}{{\cal O}}
\newcommand{\cP}{{\cal P}}

\newcommand{\e}{\mbox{\rm e}}

\newcommand{\no}{\nonumber}

\def\CR{\nonumber \\ }
\begin{document}
%\Draft
\Preprint

\title{ASPECTS OF QUANTUM CHROMODYNAMICS\Foot }

\author{Antonio Pich}

\address{Departament de F\'{\i}sica Te\`orica, IFIC,
Universitat de Val\`encia --- CSIC\\
Apt. Correus 2085, E-46071 Val\`encia, Spain\\
E-mail: Antonio.Pich@uv.es}

\maketitle
\abstracts{These lectures provide an overview of 
Quantum Chromodynamics (QCD), the $SU(3)_C$ gauge
theory of the strong interactions.
The running of the strong coupling and the associated property of
{\it Asymptotic Freedom} are analyzed.
Some selected experimental tests and the
present knowledge of $\alpha_s$ and the quark
masses are briefly summarized.
A short description of the QCD flavour symmetries and the 
{\it dynamical breaking of chiral symmetry} is also given.
A more detailed discussion can be found in standard textbooks
\cite{AH:89,MU:87,PT:84,YN:93}
and recent reviews \cite{sorrento,reviews}.
 }

%%%%%%%%%%%

\section{Quarks and Colour}
\label{sec:introduction}	

A fast look into the Particle Data Tables \cite{PDG:98}
reveals the richness and variety of the hadronic spectrum.
The large number of known mesonic and baryonic states clearly signals the
existence of a deeper level of elementary constituents of matter:
{\it quarks} \cite{GM:64}.
In fact, the messy hadronic world can be easily understood in terms
of a few constituent spin-$\frac{1}{2}$ quark {\it flavours\/}:

\vspace{0.5cm}
\begin{center}
\renewcommand{\arraystretch}{1.5} 
\begin{tabular}{|c|ccc|}  \hline
$Q=+\frac{2}{3}$ & u & c & t    \\ \hline
$Q=-\frac{1}{3}$ & d & s & b    \\ \hline 
\end{tabular} 
\renewcommand{\arraystretch}{1.2}
\end{center}
\vspace{0.5cm}

\noindent
Assuming that mesons are $M\equiv q\bar q$ 
states, while baryons have
three quark constituents, $B\equiv qqq$, one can nicely classify 
the entire hadronic spectrum.
There is a one--to--one correspondence between the
observed hadrons and the states predicted by this simple classification.
Thus, the {\it Quark Model} appears to be a very useful
{\it Periodic Table of Hadrons}.

%%%%%%%%%%%%%%%  FIGURE  %%%%%%%%%%%%%%%%%%%%%%%%
\begin{figure}[tbh]
\myfigure{\epsfxsize = 5cm \epsfbox{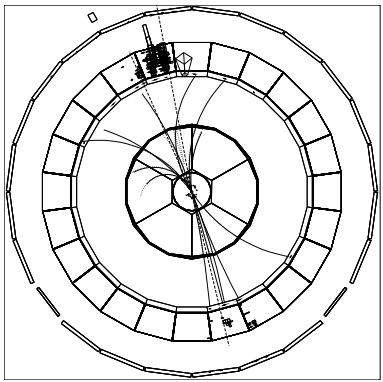}} 
\caption{Two-jet event from the hadronic decay of a $Z$ boson (DELPHI).}
\label{fig:two_jet}
\end{figure}
%%%%%%%%%%%%%%%%%%%%%%%%%%%%%%%%%%%%%%%%%%%%%%%%%

However, the quark picture faces a problem concerning the Fermi--Dirac
statistics of the constituents.
Since the fundamental state of a composite system is expected to have
$L=0$, the $\Delta^{++}$ baryon ($J=\frac{3}{2}$) 
with $J_3=+\frac{3}{2}$ corresponds to 
$\, u^\uparrow u^\uparrow u^\uparrow \, $,
with the three quark spins aligned into the same direction
($s_3=+\frac{1}{2}$) and all relative angular momenta equal to zero.
The wave function is symmetric and, therefore, the
$\Delta^{++}$ state obeys the wrong statistics.
The problem can be solved assuming \cite{GEL:72}
the existence of a new quantum
number, {\it colour}, such that each species of quark may have $N_C=3$
different colours: $q^\alpha$, $\alpha =1,2,3$ (red, yellow, violet).
Then, one can reinterpret this state as \
$\Delta^{++}\sim\epsilon^{\alpha\beta\gamma}\,
|u^\uparrow_\alpha u^\uparrow_\beta u^\uparrow_\gamma\rangle $ \
(notice that at least 3 colours are needed for making an antisymmetric 
state).
In this picture, baryons and mesons are described by
the colour--singlet combinations
\be\label{eq:m_b_wf}
B\, =\, {1\over\sqrt{6}}\,\epsilon^{\alpha\beta\gamma}\,
|q_\alpha q_\beta q_\gamma\rangle \, ,
\qquad\qquad
M\, =\, {1\over\sqrt{3}}\,\delta^{\alpha\beta} \,
|q_\alpha \bar q_\beta \rangle \, .
\ee
In order to avoid the existence of non-observed extra states
with non-zero colour,
one needs to further postulate that all asymptotic states
are colourless, \ie\ singlets under rotations in colour space.
This assumption is known as the {\it confinement hypothesis},
because it implies the non-observability of free quarks:
since quarks carry colour they are confined within
colour--singlet bound states.

The quark picture is not only a nice mathematical scheme to classify the
hadronic world. We have strong experimental evidence of the existence of
quarks. Fig. \ref{fig:two_jet} shows a typical 
$Z\to \mbox{\rm hadrons}\, $ event.
Although there are many hadrons in the final state, they
appear to be collimated in 2 {\it jets} of particles, as expected from a
two-body decay $Z\to q\bar q$, where the $q\bar q$ pair has later 
{\it hadronized}.

\subsection{Evidence of Colour}

%%%%%%%%%%%%%%%   FIGURE  %%%%%%%%%%%%%%%%%%%%%%%%%%%%%%
\begin{figure}[tbh]
\centerline{
\begin{minipage}[c]{.48\linewidth}
%\begin{center}
\mbox{}\vspace{0.5cm}
\myfigure{\epsfysize = 2cm \epsfbox{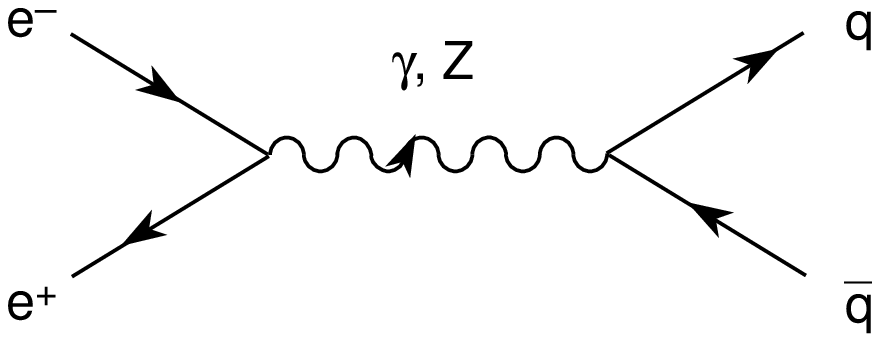} } 
\caption{$e^+e^-\to \mbox{\rm hadrons}$.}
\label{fig:eediagram}
%\end{center}
\end{minipage}
\hfill
\begin{minipage}[c]{.48\linewidth}
%\begin{center}
\myfigure{\epsfysize=2.5cm \epsfbox{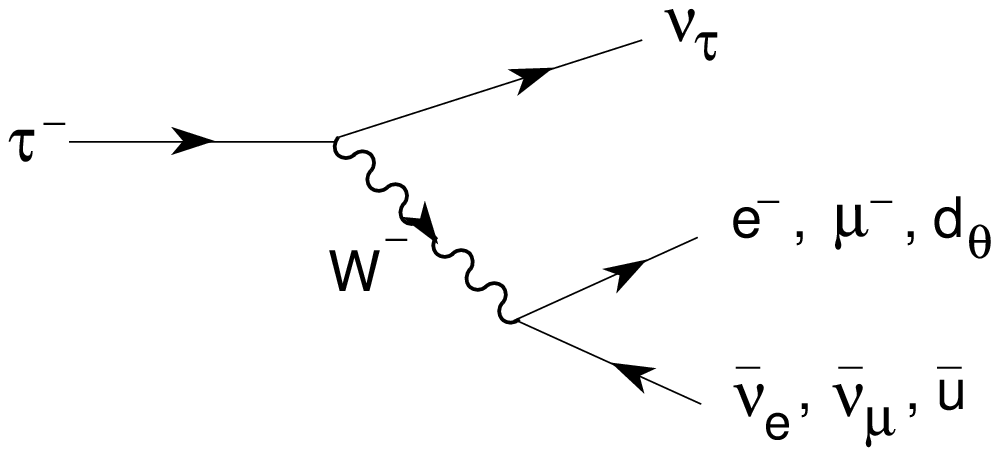}}
\caption{$\tau$ decay.}
\label{fig:tau}
%\end{center}
\end{minipage}}
\end{figure}
%%%%%%%%%%%%%%%%%%%%%%%%%%%%%%%%%%%%%%%%%%%%%%%%%%%%%%%%%

A direct test of the colour quantum number can be obtained from the 
ratio
\be\label{eq:R_ee}
R_{e^+e^-} \equiv 
{\sigma(e^+e^-\to \mbox{\rm hadrons})\over\sigma(e^+e^-\to\mu^+\mu^-)} \, .
\ee
The hadronic production occurs through
$e^+e^-\to\gamma^*, Z^*\to q\bar q\to \mbox{\rm hadrons}$. 
Since quarks are assumed
to be confined, the probability to hadronize is just one; therefore, 
summing over all possible quarks in the final state,
we can estimate the inclusive cross-section into hadrons.
At energies well below the $Z$ peak, the cross-section is 
dominated by the $\gamma$--exchange amplitude;
the ratio $R_{e^+e^-}$ is then given by the sum of the 
quark electric charges squared:
\be\label{eq:R_ee_res}
R_{e^+e^-} \approx N_C \, \sum_{f=1}^{N_f} Q_f^2 \, = \,
\left\{ 
\begin{array}{cc}
\frac{2}{3} N_C = 2\, , \qquad & (N_f=3 \; :\; u,d,s)  \\
\frac{10}{9} N_C = \frac{10}{3}\, , \qquad & (N_f=4 \; :\; u,d,s,c)  \\
\frac{11}{9} N_C = \frac{11}{3} \, ,\qquad & (N_f=5 \; :\; u,d,s,c,b)  
\ea\right. .
\ee
%

%%%%%%%%%%%%%%%  FIGURE  %%%%%%%%%%%%%%%%%%%%
\begin{figure}[tbh]
\myfigure{\epsfxsize = 10cm \epsfbox{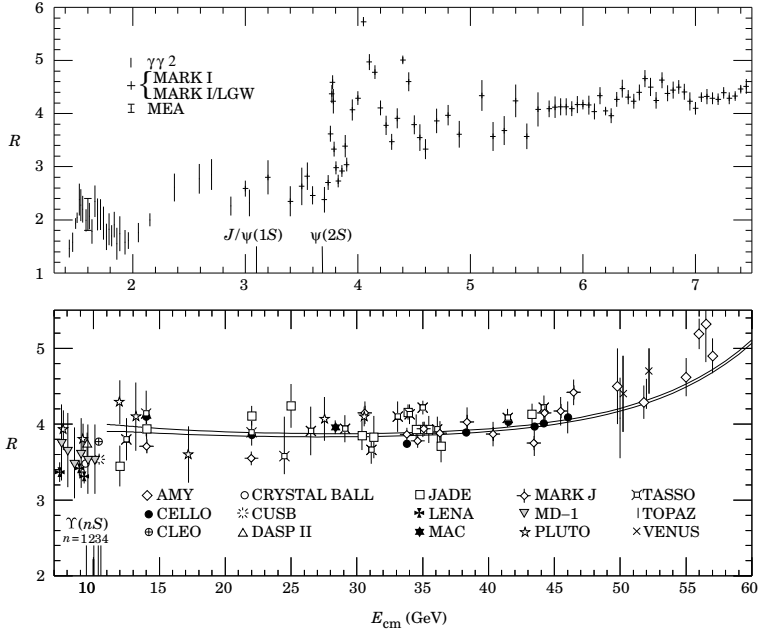}}
\caption{Measurements \protect\cite{PDG:98} of $R_{e^+e^-}$.
The two continuous curves are QCD fits.}
\label{fig:Ree}
\end{figure}
%%%%%%%%%%%%%%%%%%%%%%%%%%%%%%%%%%%%%%%%%%%%%

The measured ratio is shown in Fig. \ref{fig:Ree}. Although 
the simple
formula \eqn{eq:R_ee_res} cannot explain the complicated
structure around the different quark thresholds, 
it gives the right average
value of the cross-section (away from thresholds),
provided that $N_C$ is taken to be three.
The agreement is better at larger energies.
Notice that strong 
interactions have not been taken into account;
only the confinement hypothesis has been used.

Additional evidence for $N_C=3$ is obtained from the
hadronic decay of the $\tau$ lepton, which 
proceeds through the $W$--emission diagram
shown in Fig.~\ref{fig:tau}. Since the $W$ coupling to the charged
current is of universal strength, there are $(2+N_C)$ equal
contributions 
(if final masses and strong interactions are neglected)
to the $\tau$ decay width. Two of them correspond
to the leptonic decay modes $\tau^-\to\nu_\tau e^-\bar\nu_e$ and
$\tau^-\to\nu_\tau \mu^-\bar\nu_\mu$, while the other $N_C$
are associated with the possible colours of the quark--antiquark
pair in the $\tau^-\to\nu_\tau d_\theta u$ decay mode
($d_\theta\equiv \cos\theta_C d + \sin\theta_C s$).
Hence, the branching ratios for the different channels are expected
to be approximately:
\bel{eq:B_l}
B_{\tau\to l}\equiv 
{\rm Br}(\tau^-\to\nu_\tau l^-\bar\nu_l) 
\approx {1\over 2+N_C} = {1\over 5} = 20 \% \,  ,
\ee
\bel{eq:R_tau}
R_\tau \equiv 
{\Gamma(\tau^-\to\nu_\tau + \mbox{\rm hadrons})\over
   \Gamma(\tau^-\to\nu_\tau e^-\bar\nu_e)} \approx N_C = 3 \, ,
\ee
which should be compared with the experimental averages \cite{lp99tau}:
\beqn\label{eq:B_l_exp}
B_{\tau\to e} = (17.791\pm 0.054)\% \, , \qquad\qquad
B_{\tau\to \mu} = (17.333\pm 0.054)\% \, ,
\\ \label{eq:R_tau_exp}
R_\tau = (1-B_{\tau\to e}-B_{\tau\to \mu})/B_{\tau\to e} 
= 3.647\pm 0.014 \, . \qquad\qquad
\eeqn
The agreement is fairly good. Taking $N_C=3$, the naive predictions
only deviate from the measured values by about 20\%.
Many other observables, such as the partial widths of the $Z$ and $W^\pm$
bosons, can be analyzed in a similar way to conclude that $N_C=3$.

%%%%%%%%%%%%%%%
\begin{figure}[tb]
\myfigure{\epsfxsize = 9cm \epsfbox{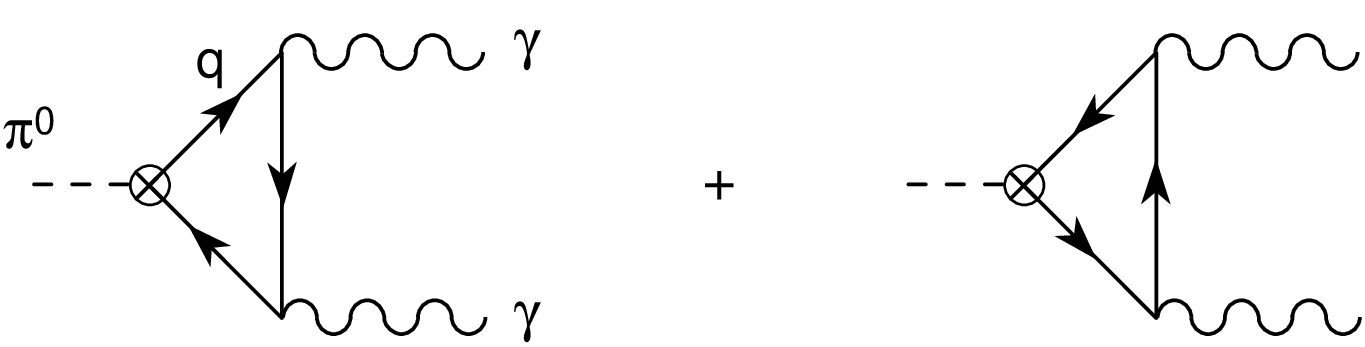}} 
\caption{Triangular quark loops generating the decay $\pi^0\to\gamma\gamma$.}
\label{fig:triangle} 
\end{figure}
%%%%%%%%%%%%%%%

A particularly strong test is obtained from the $\pi^0\to\gamma\gamma$ decay,
which occurs through the triangular quark loops in Fig.~\ref{fig:triangle}.
The crossed vertex denotes the axial current 
$A_\mu^3\equiv (\bar u \gamma_\mu\gamma_5 u - \bar d \gamma_\mu\gamma_5 d)$.
One gets:
\be\label{eq:pi_decay}
\Gamma(\pi^0\to\gamma\gamma) = \left( {N_C\over 3}\right)^2
{\alpha^2 m_\pi^3\over 64 \pi^3 f_\pi^2} = 7.73 \, {\rm eV} ,
\ee
where the $\pi^0$ coupling to $A_\mu^3$,
$f_\pi = 92.4$ MeV, is known from  the
$\pi^-\to\mu^-\bar\nu_\mu$ decay rate (assuming isospin symmetry).
The agreement with the measured value \cite{PDG:98},
$\Gamma = 7.7\pm 0.6$ eV, is remarkable.
With $N_C=1$, the prediction would have failed by a factor of 9.
The nice thing about this decay is that it is associated with a 
{\it quantum anomaly\/}: a global flavour symmetry which is broken
by quantum effects (the triangular loops). One can proof that
the decay amplitude \eqn{eq:pi_decay} does not get corrected by
strong interactions \cite{AB:69}.

Anomalies provide another compelling theoretical reason to
adopt $N_C=3$. The gauge symmetries of the Standard Model of 
electroweak interactions have also anomalies associated with
triangular fermion loops 
(diagrams of the type shown in Fig.~\ref{fig:triangle}, but with
arbitrary gauge bosons ---$W^\pm,Z,\gamma$--- in the external legs and
Standard Model fermions in the internal lines).
These gauge anomalies are deathly because
they destroy the renormalizability of the theory. Fortunately,
the sum of all possible triangular loops cancels if $N_C=3$.
Thus, with three colours, anomalies are absent and the Standard
Model is well defined.

\subsection{Asymptotic Freedom}
\label{subsec:AF}

%%%%%%%%%%%%%%%  FIGURE  %%%%%%%%%%%%%%%%%%%%%%%%%
\begin{figure}[tbh]
\myfigure{\epsfysize = 3cm \epsfbox{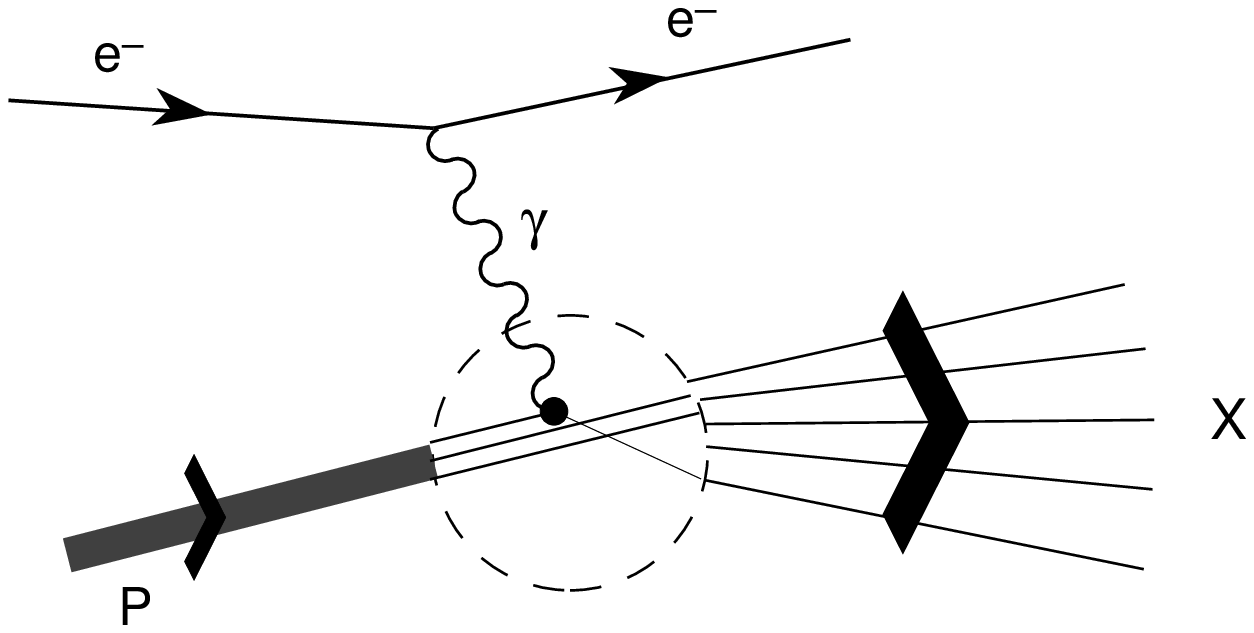}} 
\caption{Inelastic $e^-p\to e^-X$ scattering.}
\label{fig:ep_eX} 
\end{figure}
%%%%%%%%%%%%%%%%%%%%%%%%%%%%%%%%%%%%%%%%%%%%%%%%%%

The proton is an extended object with a size of the order of 1~fm.  
Its structure can be investigated through the electromagnetic 
scattering $e^- p\to e^- p$. At very low energies, the photon probe 
is unable to get information on the proton structure and
the proton behaves as a pointlike particle.
At higher energies, the photon is sensitive to shorter distances; the proton
finite size gives then rise to form factors, which suppress the elastic
cross-section at large values of
$Q^2\equiv - q^2 = 4 E E'\sin^2{\theta\over 2}$,
where $q^\mu\equiv (k_e - k'_e)^\mu$ is the momentum 
transfer through the intermediate photon propagator,
$E$ and $E'$ the energies of the incident and scattered
electrons in the proton rest frame,  and $\theta$
the scattering angle.

One can try to further resolve the proton structure, by increasing the
incident energy. The inelastic scattering
$e^- p \to e^- X$ becomes then the dominant process. Making an inclusive
sum over all hadrons produced, one has an additional kinematical variable
corresponding to the final hadronic mass, $W^2\equiv P_X^2$. The scattering
is usually described in terms of $Q^2$ and
\be\label{eq:nu}
\nu \equiv {(P\cdot q)\over M_p} = {Q^2 + W^2 - M_p^2 \over 2 M_p}
 = E - E' \, ,
\ee
where $P^\mu$ and $M_p$ are the proton cuadrimomentum and mass;
$\nu$ is the energy transfer in the proton rest frame.
In the one-photon approximation, the unpolarized differential
cross-section is given by
\be\label{eq:sigma_inel}
{d \sigma\over d Q^2 \, d \nu} =
{\pi\alpha^2\cos^2\! {\theta\over 2} \over 
4 E^2 \sin^4\! {\theta\over 2} E E'}\,
\left\{  W_2(Q^2,\nu) + 2\, W_1(Q^2,\nu)\, \tan^2{\! {\theta\over 2}} \right\} .
\ee
The proton structure is then characterized by two measurable 
{\it structure functions}. 
For a pointlike proton, the elastic scattering corresponds to
\be\label{eq:W-elastic}
W_1(Q^2,\nu) = {Q^2\over 4 M_p^2}\:\delta\!\left(\nu - {Q^2\over 2 M_p}\right) ,
\qquad\quad
W_2(Q^2,\nu) = \delta\!\left(\nu - {Q^2\over 2 M_p}\right) .
\ee

At low $Q^2$, the experimental data reveals prominent resonances;
but this resonance structure quickly dies out as $Q^2$ increases.
A much softer but sizeable continuum contribution persists at large $Q^2$,
suggesting the existence of pointlike objects inside the proton. 

To get an idea of the possible behaviour of the structure functions, one
can make a very rough model of the proton, assuming that it consist of
some number of pointlike spin-$\frac{1}{2}$ constituents (the so-called
{\it partons}), each one carrying a given fraction $\xi_i$ of the
proton momenta, \ie\ ${p_i}^\mu = \xi_i P^\mu$. That means that we are
neglecting\footnote{
%%%%%%%%%%
These approximations can be made more precise going to the
infinite momentum frame of the proton, where the transverse motion
is negligible compared with the large longitudinal boost of the partons.}
%%%%%%%%%%%%%%
the transverse parton momenta, and $m_i = \xi M_p$.
The interaction of the photon probe with the parton $i$ generates a
contribution to the structure functions given by:
\beqn\label{eq:W1}
W_1^{(i)}(Q^2,\nu) &=& {e_i^2 Q^2\over 4 m_i^2}\:
\delta\!\left(\nu - {Q^2\over 2 m_i}\right) = 
{e_i^2\over 2 M_p}\: \delta\!\left(\xi_i - x\right) ,
\CR
W_2^{(i)}(Q^2,\nu) &=& e_i^2 \:\delta\!\left(\nu - {Q^2\over 2 m_i}\right) = 
e_i^2 \, {x\over\nu}\: \delta\!\left(\xi_i - x\right) ,
\eeqn
where $e_i$ is the parton electric charge and
\bel{eq:x_def}
x\equiv {Q^2\over 2 M_p\nu} = {Q^2\over Q^2 + W^2 - M_p^2} \, .
\ee
Thus, the parton structure functions only depend on the ratio $x$,
which, moreover, fixes the momentum fractions $\xi_i$.
We can go further, and assume that in the limit $Q^2\to\infty$,
$\nu\to\infty$, but keeping $x$ fixed, the proton structure functions
can be estimated from an incoherent sum of the parton ones
(neglecting any strong interactions among the partons). Denoting 
$f_i(\xi_i)$ the probability that the parton $i$ has momentum fraction
$\xi_i$, one then has:
\beqn\label{eq:W1p}
W_1(Q^2,\nu) & =& \sum_i \int_0^1 d\xi_i\, f_i(\xi_i)\,
W_1^{(i)}(Q^2,\nu) = {1\over 2 M_p} \sum_i e_i^2\, f_i(x)
\equiv {1\over M_p}\, F_1(x) \, ,\quad
\CR
W_2(Q^2,\nu) & =&\sum_i \int_0^1 d\xi_i\, f_i(\xi_i)\,
W_2^{(i)}(Q^2,\nu) = {x\over \nu}\, \sum_i e_i^2\, f_i(x)
\equiv {1\over \nu}\, F_2(x) \, .
\eeqn
This simple parton description implies then the so-called Bjorken
 \cite{BJ:69} {\it scaling\/}: 
the proton structure functions only depend on the 
kinematical variable $x$. Moreover, one gets the Callan--Gross 
relation \cite{CG:69}
\bel{eq:CG}
F_2(x) = 2 x F_1(x) \, ,
\ee
which is a consequence of our assumption of spin-$\frac{1}{2}$ partons.
It is easy to check that spin-0 partons would have lead to $F_1(x)=0$.

%%%%%%%%%%%%%%%%%%%%%  FIGURES %%%%%%%%%%%%%%%%%%%%%%%%%%
\begin{figure}[tbh]
%\vfill
\centerline{
\begin{minipage}[t]{.48\linewidth}
\myfigure{\mbox{\epsfxsize=6cm \epsffile{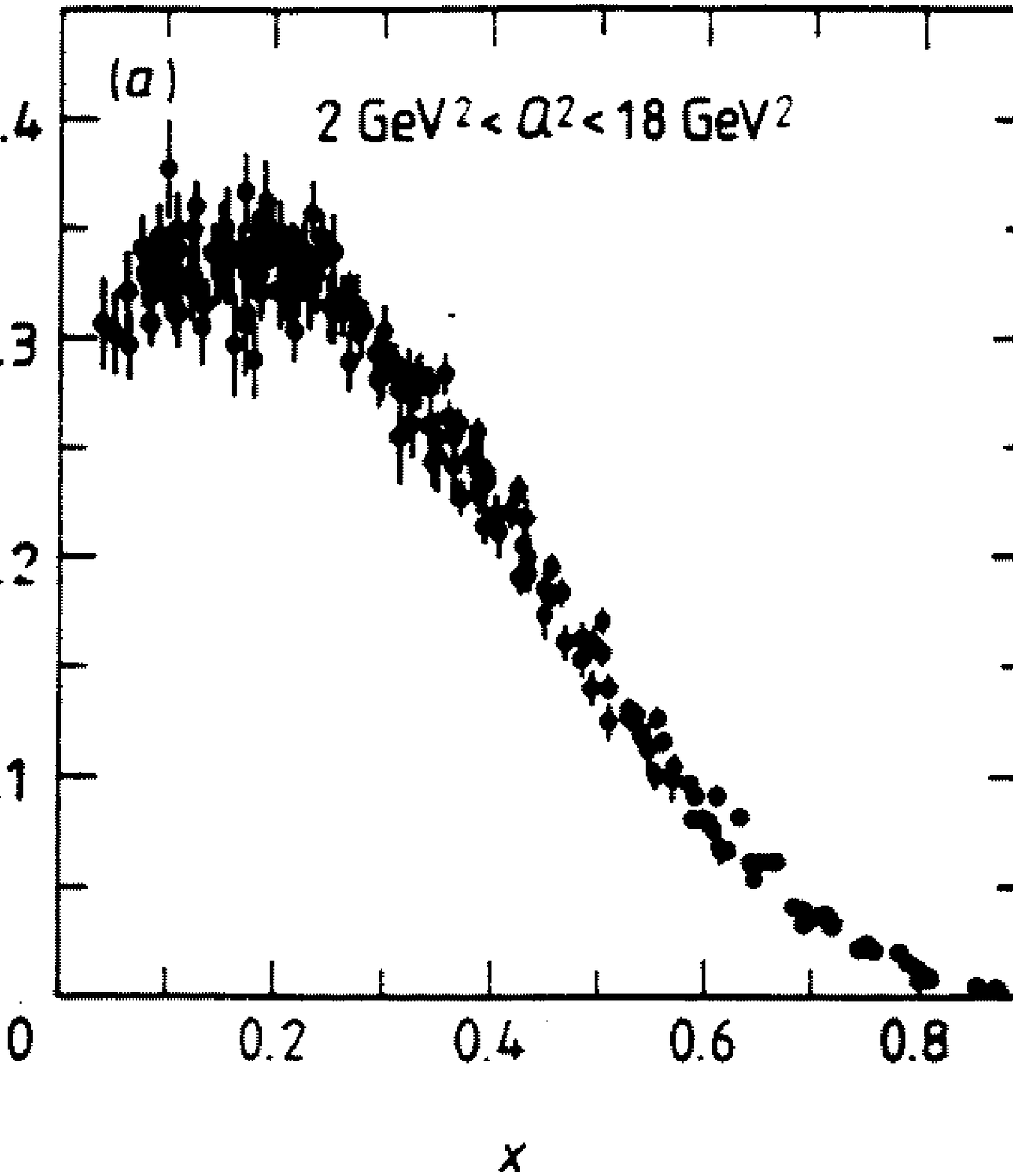}}}
%\vspace{-0.3cm}
\caption{Experimental data \protect\cite{AT:80}
on $\nu W_2$ as function of $x$, for different values of $Q^2$ 
(taken from Ref.~1).}   %\protect\citenum{AH:89}).}
\label{fig:W2x}
\end{minipage}
\hspace{.05\linewidth}
\begin{minipage}[t]{.44\linewidth}
 \myfigure{\mbox{\epsfysize=5.8cm\epsffile{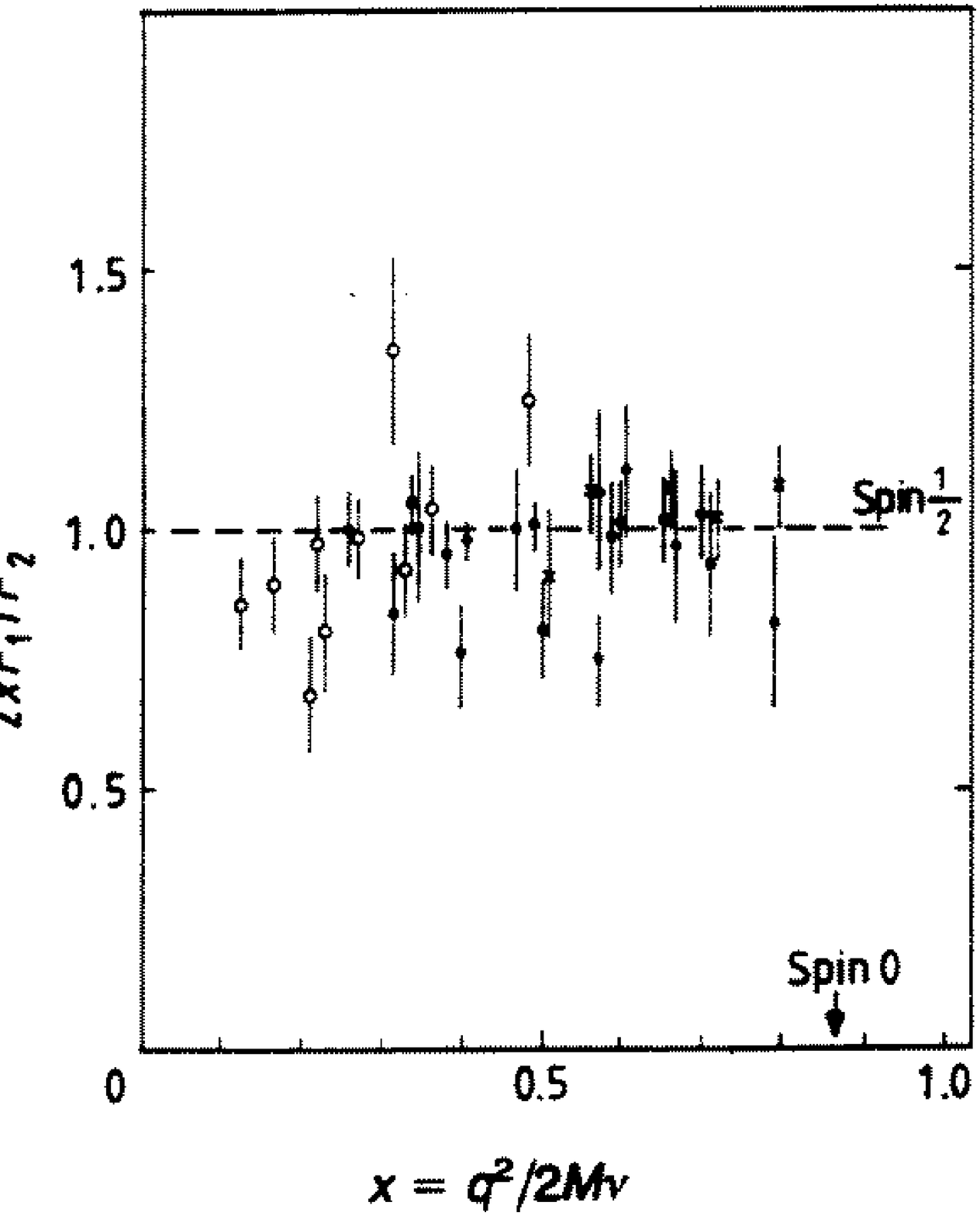}}}
%\vspace{-0.1cm}
\caption{The ratio $2xF_1/F_2$ versus $x$,
for different $Q^2$ values  \protect\cite{PE:87} 
 (1.5 GeV$^2 < Q^2 < 16$ GeV$^2$) 
(taken from Ref.~1).}   %\protect\citenum{AH:89}).}
\label{fig:CG}
\end{minipage}}
%\vfill
\end{figure}
%%%%%%%%%%%%%%%%%%%%% End figures %%%%%%%%%%%%%%%%%%%%%%%%%%

The measured values of $\nu W_2(Q^2,\nu)$ are shown in Fig.~\ref{fig:W2x}
as function of $x$, for many different values of $Q^2$ between 2 and 18
GeV$^2$; 
the concentration of data points along a curve indicates that Bjorken scaling
is correct, to a quite good approximation. 
Fig.~\ref{fig:CG} shows that the Callan--Gross relation is also
well satisfied by the data, supporting the spin-$\frac{1}{2}$ assignment
for the partons. 

The surprising thing of these successful predictions is that we have assumed
the existence of free independent pointlike partons inside the proton, in
spite of the fact that quarks are supposed to be confined by very strong
colour forces.
Bjorken scaling suggests that the strong interactions must have the
property of {\it asymptotic freedom\/}: 
they should become weaker at short distances,
so that quarks behave as free particles for $Q^2\to\infty$.
This also agrees with the empirical observation in Fig.~\ref{fig:Ree},
that the free--quark description of the ratio $R_{e^+e^-}$ works 
better at higher energies.

Thus, the interaction between a $q\bar q$ pair looks like some kind of
rubber band. If we try to separate the quark form the antiquark
the force joining them increases.
At some point, the energy on the elastic band is larger than $2m_{q'}$, so
that it becomes energetically favourable to create an additional $q'\bar q'$
pair; then the band breaks down into two mesonic systems, $q\bar q'$ and
$q'\bar q$, each one with its corresponding half-band joining the quark pair. 
Increasing more and more the energy, we can only produce more and more 
mesons, but quarks remain always confined within colour--singlet bound states.
Conversely, if one tries to approximate two quark constituents into
a very short-distance region, the elastic band loses the energy and becomes
very soft; quarks behave then as free particles.

\subsection{Why $SU(3)$?}

Flavour--changing transitions have a much weaker strength than processes
mediated by the strong force. The quark flavour is associated
with the electroweak interactions, while strong forces appear to be
flavour--conserving and flavour--independent.
On the other side, the carriers of the electroweak interaction
($\gamma$, $Z$, $W^\pm$) do not couple to the quark colour. Thus, it seems
natural to take colour as the charge associated with the strong forces 
and try to build a quantum field theory based on it \cite{FGL:73}.
The empirical evidence described so far puts a series of requirements that
the fundamental theory of colour interactions should satisfy:
\begin{enumerate}
\item Exact colour symmetry $G_C$ 
  (hadrons do not show colour multiplicity).
\item $N_C=3$. Thus, quarks belong to the triplet representation 
    $\underline{3}$ of $G_C$. 
\item Quarks and antiquarks are different states. Therefore,
  $\underline{3}^*\not= \underline{3}$.
\item Confinement hypothesis: hadronic states are colour singlets.
\item Asymptotic freedom.
\end{enumerate}

Among all compact simple Lie groups there are only four having
irreducible representations of dimension 3; 
%moreover, 
three of them are isomorphic to 
each other. Thus, we have only two choices: $SU(3)$ or
$SO(3)\simeq SU(2)\simeq Sp(1)$. Since the triplet representation of $SO(3)$
is real, only the symmetry group $SU(3)$ survives the conditions 1, 2 and 3.
The well-known $SU(3)$ decomposition of the products of $\underline{3}$
and $\underline{3}^*$ representations,
\beqn\label{eq:rep_products}
q\bar q: & & \underline{3}\otimes \underline{3}^* = 
  \underline{1} \oplus \underline{8} \, ,
\no\\
qqq: && \underline{3}\otimes\underline{3}\otimes\underline{3} =
  \underline{1} \oplus \underline{8} \oplus \underline{8}
  \oplus \underline{10} \, ,
\no\\
qq: &&  \underline{3}\otimes \underline{3} = 
  \underline{3}^* \oplus \underline{6} \, ,
\no\\
qqqq: && 
  \underline{3}\otimes\underline{3}\otimes\underline{3}\otimes \underline{3}= 
  \underline{3}\oplus\underline{3}\oplus\underline{3}\oplus\underline{6}^*
  \oplus\underline{6}^*\oplus\underline{15}\oplus\underline{15}
  \oplus\underline{15}\oplus\underline{15'} \, , 
\no\eeqn
guarantees that there are colour--singlet configurations corresponding to
meson ($q\bar q$) and baryon ($qqq$) states, as required
by the confinement hypothesis. Other exotic combinations such as diquarks
($qq$, $\bar q\bar q$) or four--quark states 
($qqqq$, $\bar q\bar q\bar q\bar q$) do not satisfy this requirement.

Clearly, the theory of colour interactions should be based on $SU(3)_C$. 
It remains to be seen whether such a theory is able
to explain confinement and asymptotic freedom as natural dynamical
consequences of the colour forces.

\setcounter{equation}{0}
\section{Gauge Symmetry: QED}
\label{sec:qed}

Let us consider the Lagrangian describing a free Dirac fermion:
\bel{eq:l_free}
\cL_0\, =\, i \,\overline{\Psi}(x)\gamma^\mu\partial_\mu\Psi(x)
\, - \, m\, \overline{\Psi}(x)\Psi(x) \, .
\ee
$\cL_0$ is invariant under {\em global} \
$U(1)$ transformations
\bel{eq:global}
\Psi(x) \;\toU\; \Psi'(x)\,\equiv\,\exp{\{i Q \theta\}}\,\Psi(x) \, ,
\ee
where $Q\theta$ is an arbitrary real constant.
The phase of $\Psi(x)$ is then a pure convention--dependent
quantity without physical meaning. 
However,
the free Lagrangian is no-longer invariant if one allows
the phase transformation to depend on the space--time coordinate,
\ie\ under {\em local} phase redefinitions $\theta=\theta(x)$,
because
\bel{eq:local}
\partial_\mu\Psi(x) \;\toU\; \exp{\{i Q \theta\}}\,
\left(\partial_\mu + i Q \partial_\mu\theta\right)\,
\Psi(x) \, .
\ee
Thus, once an observer situated at the reference point $x_0$
has adopted a given phase convention, the same convention must
be taken at all space--time points. This looks very unnatural.

The ``Gauge Principle'' is the requirement that the $U(1)$
phase invariance should hold {\em locally}.
This is only possible if one adds some additional piece to the
Lagrangian, transforming in such a way  as to cancel the 
$\partial_\mu\theta$ term in Eq.~\eqn{eq:local}.
The needed modification is completely fixed by the transformation 
\eqn{eq:local}: one introduces a new spin-1 
(since $\partial_\mu\theta$  has a Lorentz index)
field $A_\mu(x)$, transforming as
\bel{eq:a_transf}
A_\mu(x)\;\toU\; A_\mu'(x)\,\equiv\, A_\mu(x) + {1\over e}\,
\partial_\mu\theta\, ,
\ee
and defines the covariant derivative
\bel{eq:d_covariant}
D_\mu\Psi(x)\,\equiv\,\left[\partial_\mu-ieQA_\mu(x)\right]
\,\Psi(x)\, ,
\ee
which has the required property of transforming like the field itself:
\bel{eq:d_transf}
D_\mu\Psi(x)\;\toU\;\left(D_\mu\Psi\right)'(x)\,\equiv\,
\exp{\{i Q \theta\}}\,D_\mu\Psi(x)\,.
\ee
The Lagrangian
\bel{eq:l_new}
\cL\,\equiv\, 
i \,\overline{\Psi}(x)\gamma^\mu D_\mu\Psi(x)
\, - \, m\, \overline{\Psi}(x)\Psi(x) 
\, =\, \cL_0\, +\, e Q A_\mu(x)\, \overline{\Psi}(x)\gamma^\mu\Psi(x)
\ee
is then invariant under local $U(1)$ transformations.

The gauge principle has generated an interaction 
between the Dirac spinor and the gauge field $A_\mu$,
which is nothing else than the familiar vertex of
Quantum Electrodynamics (QED).
Note that the corresponding electromagnetic
charge $e Q$ is completely arbitrary.
If one wants $A_\mu$ to be a true propagating field, one needs to add
a gauge--invariant kinetic term
\bel{eq:l_kinetic}
\cL_{\mbox{\rms Kin}}\,\equiv\, -{1\over 4} F_{\mu\nu} F^{\mu\nu}\,,
\ee
where
$F_{\mu\nu}\,\equiv\, \partial_\mu A_\nu -\partial_\nu A_\mu$
is the usual electromagnetic field strength.
A possible mass term for the gauge field, 
${1\over 2}m^2A^\mu A_\mu$, is forbidden because it would violate
gauge invariance; therefore, 
the photon field is predicted to be massless.
The total Lagrangian in \eqn{eq:l_new} and \eqn{eq:l_kinetic}
gives rise to the well-known Maxwell equations.

From a simple gauge--symmetry requirement, we have deduced
the right QED Lagrangian, which leads to a very  successful
quantum field theory. 
Remember that QED predictions have been tested
to a very high accuracy.

\setcounter{equation}{0}
\section{The QCD Lagrangian}
\label{sec:QCDlagrangian}

Let us denote $q^\alpha_f$ a quark field of colour $\alpha$ and flavour $f$.
To simplify the equations, let us adopt a vector notation
in colour space:
$q_f^T \,\equiv\, (q^1_f,q^2_f,q^3_f) $.
The free Lagrangian
\bel{eq:L_free}
\cL_0 = \sum_f\,\bar q_f \, \left( i\gamma^\mu\partial_\mu - m_f\right) q_f
\ee
is invariant under arbitrary global $SU(3)_C$ transformations in colour
space,
\bel{eq:q_transf}
q^\alpha_f \,\longrightarrow\, 
(q^\alpha_f)' = U^\alpha_{\phantom{\alpha}\beta}\, q^\beta_f \, ,
\qquad\quad
U U^\dagger = U^\dagger U = 1 \, , \qquad\quad
\det U = 1 \, .
\ee
The $SU(3)_C$ matrices can be written in the form
\bel{eq:U_def}
U = \exp\left\{-ig_s {\lambda^a\over 2}\theta_a\right\} \, ,
\ee
where $\lambda^a$ ($a=1,2,\ldots,8$) denote the
generators of the fundamental representation of the
$SU(3)_C$ algebra, 
and $\theta_a$ are arbitrary parameters. The matrices $\lambda^a$
are traceless and
satisfy the commutation relations
\bel{eq:commutation}
%\left[{\lambda^a\over 2},{\lambda^b\over 2}\right] = 
%i f^{abc} \, {\lambda^c\over 2} \, ,
\left[\lambda^a,\lambda^b\right] = 
2 i f^{abc} \, \lambda^c \, ,
\ee
with $f^{abc}$ the $SU(3)_C$
structure constants, which are real and totally antisymmetric.
Some useful properties of $SU(3)$ matrices are collected in Appendix~A.

As in the QED case, we can now require the Lagrangian to be also invariant
under {\it local} $SU(3)_C$ transformations, $\theta_a = \theta_a(x)$. To
satisfy this requirement, we need to change the quark derivatives by covariant
objects. Since we have now 8 independent gauge parameters, 8 different
gauge bosons $G^\mu_a(x)$, the so-called {\it gluons}, are needed:
\bel{eq:D_cov}
D^\mu q_f \,\equiv\, \left[ \partial^\mu - i g_s {\lambda^a\over 2} 
G^\mu_a(x)\right]
  \, q_f \, \equiv\, \left[ \partial^\mu - i g_s G^\mu(x)\right] \, q_f \, .
\ee
Notice that we have introduced the compact matrix notation
\bel{eq:G_matrix}
  [G^\mu(x)]_{\alpha\beta}\,\equiv\, 
\left({\lambda^a\over 2}\right)_{\!\alpha\beta}\, G^\mu_a(x) \, .
\ee
We want $D^\mu q_f$ to transform in exactly the same way as the 
colour--vector $q_f$;
this fixes the transformation properties of the gauge fields:
$$  %\bel{eq:G_trans}
D^\mu \,\longrightarrow\, (D^\mu)'= U\, D^\mu\, U^\dagger \, ; \qquad
G^\mu \,\longrightarrow\, (G^\mu)'= U\, G^\mu\, U^\dagger 
  -{i\over g_s} \, (\partial^\mu U) \, U^\dagger \, .
$$ %\ee
Under an infinitesimal $SU(3)_C$ transformation,
\beqn\label{eq:inf_transf}
q^\alpha_f &\:\longrightarrow\: & (q^\alpha_f)' = 
q^\alpha_f -i g_s \left({\lambda^a\over 2}\right)_{\!\alpha\beta} 
\delta\theta_a\, q^\beta_f \, ,
\no\\
G^\mu_a &\:\longrightarrow\: & (G^\mu_a)' = 
G^\mu_a -\partial^\mu(\delta\theta_a)
 + g_s f^{abc} \delta\theta_b \, G^\mu_c \, .
\eeqn
The gauge transformation of the gluon fields is more complicated that the one
obtained in QED for the photon.
The non-commutativity of the $SU(3)_C$ matrices gives rise to an additional term
involving the gluon fields themselves. 
For constant $\delta\theta_a$, the transformation rule for the gauge fields
is expressed in terms of the structure constants $f^{abc}$; thus, 
the gluon fields belong to the adjoint representation of the colour group
(see Appendix~A).
Note also that there is a unique $SU(3)_C$
coupling $g_s$. In QED it was possible to assign arbitrary electromagnetic
charges to the
different fermions. Since the commutation relation \eqn{eq:commutation} is
non-linear, this freedom does not exist for $SU(3)_C$.

To build a gauge--invariant kinetic term for the gluon fields, we introduce
the corresponding field strengths:
\beqn\label{eq:G_tensor}
G^{\mu\nu}(x) & \equiv & {i\over g_s}\, [D^\mu, D^\nu] \, = \,
  \partial^\mu G^\nu - \partial^\nu G^\mu - i g_s\, [G^\mu, G^\nu] \, \equiv \,
  {\lambda^a\over 2}\, G^{\mu\nu}_a(x) \, ,
\no\\
G^{\mu\nu}_a(x) & = & \partial^\mu G^\nu_a - \partial^\nu G^\mu_a
  + g_s f^{abc} G^\mu_b G^\nu_c \, .
\eeqn
Under a gauge transformation,
\bel{eq:G_tensor_transf} 
G^{\mu\nu}\,\longrightarrow\, (G^{\mu\nu})' = U\, G^{\mu\nu}\, U^\dagger \, ,
\ee
and the colour trace
Tr$(G^{\mu\nu}G_{\mu\nu}) = \frac{1}{2} G^{\mu\nu}_aG_{\mu\nu}^a$
remains invariant.

Taking the proper normalization for the gluon kinetic term, we finally have
the $SU(3)_C$ invariant QCD Lagrangian:
\bel{eq:L_QCD}
\cL_{\mbox{\rms QCD}} \,\equiv\, -{1\over 4}\, G^{\mu\nu}_aG_{\mu\nu}^a
+ \sum_f\,\bar q_f \, \left( i\gamma^\mu D_\mu - m_f\right)\, q_f \, .
\ee
It is worth while to decompose the Lagrangian into its different pieces:
\beqn\label{eq:L_QCD_pieces}
\cL_{\mbox{\rms QCD}} & = &
 -{1\over 4}\, (\partial^\mu G^\nu_a - \partial^\nu G^\mu_a)
  (\partial_\mu G_\nu^a - \partial_\nu G_\mu^a)
 + \sum_f\,\bar q^\alpha_f \, \left( i\gamma^\mu\partial_\mu - m_f\right)\, 
q^\alpha_f
\qquad\no\\ && \mbox{}
 + g_s\, G^\mu_a\,\sum_f\,  \bar q^\alpha_f \gamma_\mu
 \left({\lambda^a\over 2}\right)_{\!\alpha\beta} q^\beta_f 
\\ && \mbox{}
 - {g_s\over 2}\, f^{abc}\, (\partial^\mu G^\nu_a - \partial^\nu G^\mu_a) \,
  G_\mu^b G_\nu^c \, - \, 
{g_s^2\over 4} \, f^{abc} f_{ade} \, G^\mu_b G^\nu_c G_\mu^d G_\nu^e \, .
\no
\eeqn
The first line contains the correct kinetic terms for the different fields, 
which
give rise to the corresponding propagators. The colour interaction between
quarks and gluons is given by the second line; it involves the $SU(3)_C$ 
matrices
$\lambda^a$. Finally, owing to the non-abelian character of the colour group,
the $G^{\mu\nu}_aG_{\mu\nu}^a$ term generates the cubic and quartic gluon
self-interactions shown in the last line; 
the strength of these interactions is given by the same coupling $g_s$ which
appears in the fermionic piece of the Lagrangian.

In spite of the rich physics contained in it, the Lagrangian \eqn{eq:L_QCD}
looks very simple because of its colour symmetry properties.
All interactions are given in terms of a single universal coupling $g_s$,
which is called the {\it strong coupling constant}.
The existence of self-interactions among the gauge fields is a new feature
that was not present in QED; it seems then reasonable to expect that
these gauge self-interactions could explain properties 
like asymptotic freedom and confinement, which do not appear in QED.

%%%%%%%%%%%%%%%  FIGURE %%%%%%%%%%%%%%%%%%%%%%%%%
\begin{figure}[tbh]
\myfigure{\epsfxsize = 5cm \epsfbox{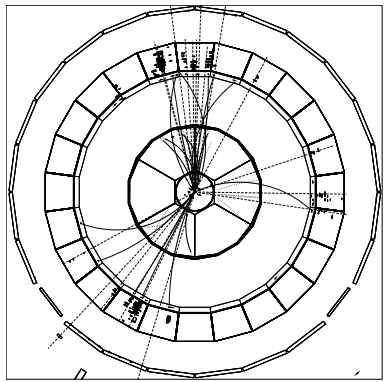}} 
\caption{Three-jet event from the hadronic decay of a $Z$ boson (DELPHI).}
\label{fig:ThreeJets}
\end{figure}
%%%%%%%%%%%%%%%%%%%%%%%%%%%%%%%%%%%%%%%%%%%%%

Without any detailed calculation, one can already extract qualitative physical
consequences from $\cL_{\mbox{\rms QCD}}$. 
Quarks can emit gluons. At lowest order
in $g_s$, the dominant process will be the emission of a single gauge boson.
Thus, the hadronic decay of the $Z$ should result in some $Z\to q\bar q G$ 
events,
in addition to the dominant $Z\to q\bar q$ decays discussed 
before.
%in Section~\ref{sec:introduction}.
Fig.~\ref{fig:ThreeJets} clearly shows that 3-jet events, with the required
kinematics, indeed appear in the LEP data. Similar events show up in
$e^+e^-$ annihilation into hadrons, away from the $Z$ peak.

In order to properly quantize the QCD Lagrangian, one needs to add 
to $\cL_{\mbox{\rms QCD}}$ the
so-called {\it Gauge-fixing} and {\it Faddeev--Popov} terms.
Since this is a rather technical issue, its discussion is relegated
to Appendix B.

\setcounter{equation}{0}
\section{Quantum Loops}
\label{sec:loops}

The QCD Lagrangian is rather economic in the sense that it involves a single
coupling $g_s$. Thus, all strong--interacting phenomena should be described 
in terms of just one parameter. At lowest order in $g_s$ (tree level), 
it is straightforward
to compute scattering amplitudes involving quarks and gluons.
%$q\bar q\to GG$, $qq\to qq$, $G q\to G q$, \ldots
Unfortunately, this exercise by itself does not help very much to understand 
the physical hadronic world.
First, we see hadrons instead of quarks and gluons. Second, we have learnt from
experiment that the strength of the strong forces changes with the
energy scale: the interaction is very strong (confining) at low energies, but
quarks behave as nearly free particles at high energies.
Obviously, we cannot understand both energy regimes with a single constant 
$g_s$,
which is the same everywhere. Moreover, if we neglect the quark masses,
the QCD Lagrangian does not contain any energy scale;
thus, there is no way to decide when the energy
of a given process is large or small, because we do not have any
reference scale to compare with.

If QCD is the right theory of the strong interactions, it should provide
some dynamical scale through quantum effects.

\subsection{Regularization of Loop Integrals}
\label{sub:regularization}

The computation of perturbative corrections to the tree-level results
involves divergent loop integrals. It is then necessary to find a way of
getting finite results with physical meaning from a priori meaningless
divergent quantities.

%%%%%%%%%%%%%%% FIGURE %%%%%%%%%%%%%%%%%%%%%
\begin{figure}[tbh] 
\myfigure{\epsfysize=2.5cm \epsfbox{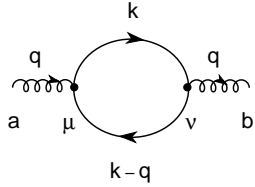}} 
\vspace{-0.2cm}
\caption{Gluon self-energy diagram.}
\label{fig:gluonSE}
\end{figure}
%%%%%%%%%%%%%%%%%%%%%%%%%%%%%%%%%%%%%%%%%%%%%%

Let us consider the self-energy gluon loop in Fig.~\ref{fig:gluonSE}.
The corresponding contribution in momentum space can be easily obtained,
using standard Feynman rules techniques: 
\bel{eq:gluon_se}
i \,\Pi^{\mu\nu}_{ab}(q) \, = \, - g_s^2 \delta_{ab} T_F \int
{d^4k\over (2\pi)^4} 
{\mbox{\rm Tr}[\gamma^\mu\slashchar{k}\gamma^\nu (\slashchar{k}-\slashchar{q})]
\over k^2 (k-q)^2} \, .
\ee
The result is proportional to $g_s^2$, because there are two $q\bar q G$ 
vertices, and there is a $SU(3)_C$ factor, $T_F = {1\over 2}$,
coming from the colour trace
${1\over 4}\mbox{\rm Tr}(\lambda^a\lambda^b) = \delta^{ab} T_F$.

The problem appears in the momentum integration, which is clearly divergent
[$\sim \int d^4k (1/k^2) = \infty$].
We can define the momentum integral in many different (and arbitrary) ways.
For instance, we could introduce a {\it cut-off} $M$, such that only
momentum scales smaller than $M$ are integrated; obviously, the resulting
integral would be an increasing function of $M$.
Instead, it has become conventional to define the loop integrals through
{\it dimensional regularization}: 
the calculation is performed in
$D=4+2\epsilon$ dimensions  \cite{dimreg}. 
For $\epsilon\not=0$ the resulting integral is
well defined:
$$ %\bel{eq:dimensional}
\int {d^Dk\over (2\pi)^D}\, {k^\alpha (k-q)^\beta \over k^2 (k-q)^2} \, = \,
{-i\over 6 (4\pi)^2} \left({-q^2\over 4\pi}\right)^\epsilon
\Gamma(-\epsilon) \, \left(1-{5\over 3}\epsilon\right)
\left\{ {q^2 g^{\alpha\beta}\over 2 (1+\epsilon)} + q^\alpha q^\beta\right\} 
 .
$$  %\ee
The ultraviolet divergence of the loop appears at $\epsilon=0$,
through the pole of the Gamma function,
\bel{eq:gammaE}
\Gamma(-\epsilon) \, = \, -{1\over\epsilon} - \gamma_E + \cO(\epsilon^2) \, ,
\qquad\qquad
\gamma_E = 0.577215\ldots
\ee

Dimensional regularization has many advantages because does not spoil 
the gauge symmetry of QCD and, therefore, simplifies a lot the calculations.
One could argue that a cut-off procedure would
be more {\it physical}, since the parameter $M$ could be related to some
unknown physics at very short distances. However, within the
QCD framework, both prescriptions are equally meaningless. One just introduces
a regularizing parameter, such that the integral is well defined and the
divergence is recovered in some limit ($M\to\infty$ or $\epsilon\to 0$).

Since the momentum--transfer $q^2$ has dimensions, 
it turns out to be convenient
to introduce an arbitrary energy scale $\mu$ and write
$$  %\bel{eq:mu_scale}
%\left({-q^2\over 4\pi}\right)^\epsilon\Gamma(-\epsilon) =
\mu^{2\epsilon} \left({-q^2\over 4\pi\mu^2}\right)^\epsilon\Gamma(-\epsilon) =
-\mu^{2\epsilon}\left\{{1\over\epsilon} +\gamma_E -\ln{4\pi}
+\ln{(-q^2/\mu^2)} + \cO(\epsilon)\right\} \, .
$$  %\ee
Obviously, this expression does not depend on $\mu$; but written in this form
one has a dimensionless quantity ($-q^2/\mu^2$) inside the logarithm.

The contribution of the loop diagram in Fig.~\ref{fig:gluonSE} 
can finally be written as
\bel{eq:g_se}
\Pi^{\mu\nu}_{ab}   =   
\delta_{ab}\,\left( -q^2 g^{\mu\nu} + q^\mu q^\nu\right)
\, \Pi(q^2) \, ,
\ee
with
$$
\Pi(q^2)  = 
 -{4\over 3} T_F \,\left({g_s\mu^\epsilon\over 4\pi}\right)^2
\left\{{1\over\epsilon} +\gamma_E -\ln{4\pi}
+\ln{(-q^2/\mu^2)} - {5\over 3} + \cO(\epsilon)\right\} \, .
$$
Owing to the ultraviolet divergence, this equation does not determine the
wanted self-energy contribution. Nevertheless, it does show how this effect
changes with the energy scale. 
If one could fix the value of
$\Pi(q^2)$ at some reference momentum transfer $q_0^2$, the result
would be known at any other scale:
\bel{eq:pi_1_2}
\Pi(q^2) = \Pi(q_0^2) - {4\over 3} T_F \, \left({g_s\over 4\pi}\right)^2
\,\ln{(q^2/q_0^2)} \, .
\ee

 We can split the self-energy contribution into a meaningless divergent piece
and a finite term, which includes the $q^2$ dependence,
\bel{eq:Pi_splitting}
\Pi(q^2) \,\equiv\, \Delta\Pi_\epsilon(\mu^2) + \Pi_R(q^2/\mu^2)\, .
\ee
This separation is ambiguous, because the finite $q^2$--independent
contributions can be splitted in many different ways. A given choice defines
a {\it scheme\/}:
\beqn\label{eq:Pi_epsilon}
\Delta\Pi_\epsilon(\mu^2) &\, =\, &\left\{ 
\begin{array}{ll}
 -{T_F\over 3\pi}\, {g_s^2\over 4\pi} \,\mu^{2\epsilon} \left[ {1\over\epsilon}
        + \gamma_E -\ln(4\pi) -{5\over 3}\right]  \quad\qquad\qquad &
     \;\;   (\mu\;\mbox{\rm scheme}) , \\
 -{T_F\over 3\pi}\, {g_s^2\over 4\pi} \,\mu^{2\epsilon}\, {1\over\epsilon} 
    \qquad &    (\mbox{\rm MS}\;\mbox{\rm scheme}) , \; \\
 -{T_F\over 3\pi}\, {g_s^2\over 4\pi} \,\mu^{2\epsilon} \left[ {1\over\epsilon}
        + \gamma_E -\ln(4\pi)\right]  \qquad & 
       (\overline{\mbox{\rm MS}}\;\mbox{\rm scheme}) , 
  \ea\right.
\CR &&\\
\Pi_R(q^2/\mu^2) &\, =\, &\left\{ 
  \begin{array}{ll}
   -{T_F\over 3\pi}\, {g_s^2\over 4\pi}\, \ln{(-q^2/\mu^2)} \qquad &
    \;\;   (\mu\;\mbox{\rm scheme}) , \\
   -{T_F\over 3\pi}\, {g_s^2\over 4\pi} \,\left[ \ln{(-q^2/\mu^2)}
        + \gamma_E -\ln(4\pi) -{5\over 3}\right]  \qquad & 
        (\mbox{\rm MS}\;\mbox{\rm scheme}) , \quad\\
    -{T_F\over 3\pi}\, {g_s^2\over 4\pi} \,\left[ \ln{(-q^2/\mu^2)}
        -{5\over 3}\right]  \qquad & (\overline{\mbox{\rm MS}}\;
    \mbox{\rm scheme}) . 
   \ea\right.
\no\eeqn
In the $\mu$ scheme, one uses the value of $\Pi(-\mu^2)$ to define the
divergent part.
MS and $\overline{\mbox{\rm MS}}$ stand for minimal subtraction 
\cite{THO:73} and
modified minimal subtraction schemes \cite{BBDM:78}; 
in the MS case, one subtracts only
the divergent $1/\epsilon$ term, while the $\overline{\mbox{\rm MS}}$
scheme puts also the $\gamma_E-\ln(4\pi)$ factor into the
divergent part.
Notice that the logarithmic $q^2$ dependence is always the same.

\subsection{Renormalization: QED}
\label{subsec:renormalization}

%%%%%%%%%%%%%%% FIGURE %%%%%%%%%%%%%%%%%%%%%%%
\begin{figure}[htb] 
\myfigure{\epsfysize=3cm \epsfbox{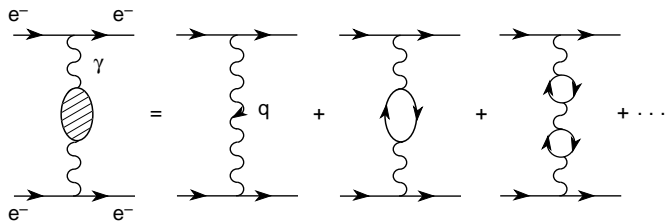}}
\caption{Photon self-energy contribution to $e^-e^-$ scattering.}
\label{fig:EEint}
\end{figure}
%%%%%%%%%%%%%%%%%%%%%%%%%%%%%%%%%%%%%%%%%%%

A Quantum Field Theory is called {\it renormalizable} if all ultraviolet
divergences can be reabsorbed through a redefinition of the original fields
and couplings.

Let us consider the electromagnetic interaction between two electrons.
At one loop, 
the QED photon self-energy contribution is just given by
Eq.~\eqn{eq:g_se}, with the changes $T_F\to 1$ and $g_s\to e$. The
corresponding scattering amplitude takes the form
\bel{eq:T_ee}
T(q^2) \,\sim\, -J^\mu J_\mu \:
 {e^2\over q^2} \, \left\{ 1 -\Pi(q^2) + \ldots
\right\} \, ,
\ee
where $J^\mu$ denotes the electromagnetic fermion current.

At lowest order, 
$T(q^2)\sim \alpha/q^2$ with $\alpha = e^2/(4\pi)$. 
The divergent correction generated by quantum loops can be reabsorbed into a
redefinition of the coupling:
\bel{eq:alpha_R}
{\alpha_0\over q^2}\, \left\{ 1 -\Delta\Pi_\epsilon(\mu^2)
- \Pi_R(q^2/\mu^2)\right\} \,\equiv\,
{\alpha_R(\mu^2)\over q^2}\, \left\{ 1 
- \Pi_R(q^2/\mu^2)\right\} \, ,
\ee
\bel{eq:alpha_Rb}
\alpha_R(\mu^2) \, = \, \alpha_0 \,
\left\{ 1 + {\alpha_0 \over 3\pi} \mu^{2\epsilon}
\left[{1\over\epsilon} + C_{\mbox{\rms scheme}}\right] + \ldots
\right\} \, ,
\qquad\quad  
\alpha_0\,\equiv\, {e_0^2\over 4\pi}\ , \quad
\ee
where $e_0$ denotes the {\it bare}
coupling appearing in the QED Lagrangian;
this bare quantity is, however, not directly observable. Making the redefinition
\eqn{eq:alpha_R}, the scattering amplitude is finite and gives rise
to a definite prediction for the cross-section, which can be compared with
experiment; thus, one actually measures the {\it renormalized} coupling
$\alpha_R$.

The redefinition \eqn{eq:alpha_R} is meaningful, provided that it can be done
in a self-consistent way: all ultraviolet divergent contributions to all
possible scattering processes should be eliminated through the same redefinition
of the coupling (and the fields). The nice thing of gauge theories, such as
QED or QCD, is that the underlying gauge symmetry guarantees the
renormalizability of the quantum field theory.

The renormalized coupling $\alpha_R(\mu^2)$ depends
on the arbitrary scale $\mu$ and on the chosen {\it renormalization scheme}
[the constant $C_{\mbox{\rms scheme}}$ denotes the different finite terms in
Eq.~\eqn{eq:Pi_epsilon}]. Quantum loops have introduced a scale dependence in a
quite subtle way. Both $\alpha_R(\mu^2)$ and the renormalized self-energy
correction $\Pi_R(q^2/\mu^2)$ depend on $\mu$, but the physical scattering
amplitude $T(q^2)$ is of course $\mu$ independent: ($Q^2\equiv -q^2$)
\beqn\label{eq:mu_dep}
T(q^2) &\sim & -4\pi\, J^\mu J_\mu\:
{\alpha_R(\mu^2)\over q^2}\, \left\{ 1 +
{\alpha_R(\mu^2)\over 3\pi} \left[ \ln{\left(-q^2\over\mu^2\right)} +
C'_{\mbox{\rms scheme}}\right] + \ldots  \right\}\no\\
& = & 4\pi\, J^\mu J_\mu\:
{\alpha_R(Q^2)\over Q^2} \, \left\{ 1 +
{\alpha_R(Q^2)\over 3\pi} C'_{\mbox{\rms scheme}} + \cdots \right\}\, .
\eeqn

The quantity $\alpha(Q^2)\equiv\alpha_R(Q^2)$ is called the QED
{\it running coupling}.
The usual fine structure constant $\alpha=1/137$ is defined through the
classical Thomson formula; therefore, it corresponds to a very low scale
$Q^2= -m_e^2$. The value of $\alpha$ relevant for LEP experiments
is not the same:
$\alpha(M_Z^2)_{\overline{\mbox{\rms MS}}} = 1/129$.
The scale dependence of $\alpha(Q^2)$ is regulated by 
the so-called $\beta$ function:
\bel{eq:beta}
\mu\, {d\alpha\over d\mu} \,\equiv\, \alpha \,\beta(\alpha) \, ;
\qquad\qquad \beta(\alpha)\, =\, \beta_1 {\alpha\over \pi} +
\beta_2 \left({\alpha\over\pi}\right)^2 + \cdots
\ee
At the one-loop level, $\beta$ reduces to its first coefficient,
which is fixed by \eqn{eq:alpha_Rb}:
\bel{eq:beta_1}
\beta_1^{\mbox{\rms QED}} \, = \, {2\over 3}\, . 
\ee
The differential equation \eqn{eq:beta} can then be easily solved,
with the result:
\bel{eq:alpha_run}
\alpha(Q^2) \, = \, {\alpha(Q_0^2)\over 1 - {\beta_1 \alpha(Q_0^2)\over 2\pi}
\ln{(Q^2/Q_0^2)}} \, .
\ee
%
%%%%%%%%%%%%%%% FIGURE %%%%%%%%%%%%%%%%%%%%%
\begin{figure}[tbh] 
\myfigure{\epsfysize=3cm \epsfbox{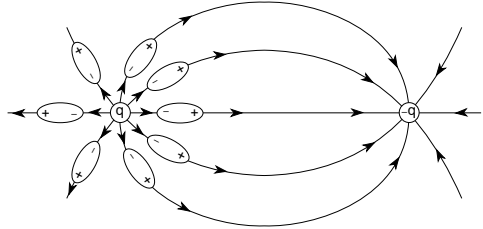}} 
\caption{Electromagnetic charge screening in a dipolar medium.}
\label{fig:screening}
\end{figure}
%%%%%%%%%%%%%%%%%%%%%%%%%%%%%%%%%%%%%%%%%%%%%%
%
Since $\beta_1>0$, the QED running coupling increases with the energy scale:
$\alpha(Q^2)>\alpha(Q_0^2)$ if $Q^2>Q_0^2$;
\ie\ the electromagnetic charge decreases at large distances. This can be
intuitively understood as the screening due to virtual $e^+e^-$ pairs
generated, through quantum effects, around the electron charge.
The physical QED vacuum behaves as a polarized dielectric medium.

Notice that taking $\mu^2=Q^2$ in Eq.~\eqn{eq:mu_dep} we have eliminated
all dependences on $\ln{(Q^2/\mu^2)}$ to all orders in $\alpha$.
The running coupling \eqn{eq:alpha_run} makes a resummation of all
leading logarithmic corrections, \ie
\bel{eq:alpha_logs}
\alpha(Q^2) \, = \, \alpha(\mu^2)\,\sum_{n=0}^\infty
\left[ {\beta_1 \alpha(\mu^2)\over 2\pi}
\ln{(Q^2/\mu^2)}\right]^n \, .
\ee
These higher-order logarithms correspond to the contributions from an
arbitrary number of one-loop self-energy insertions along the intermediate
photon propagator in Fig.~\ref{fig:EEint} \
$[1 - \Pi_R(q^2/\mu^2) + \left(\Pi_R(q^2/\mu^2)\right)^2 + \cdots]$.

\subsection{The QCD Running Coupling}
\label{subsec:AlphaRunning}

%%%%%%%%%%%%%%% FIGURE %%%%%%%%%%%%%%%%%%%%%
\begin{figure}[tbh] 
\myfigure{\epsfxsize=10cm \epsfbox{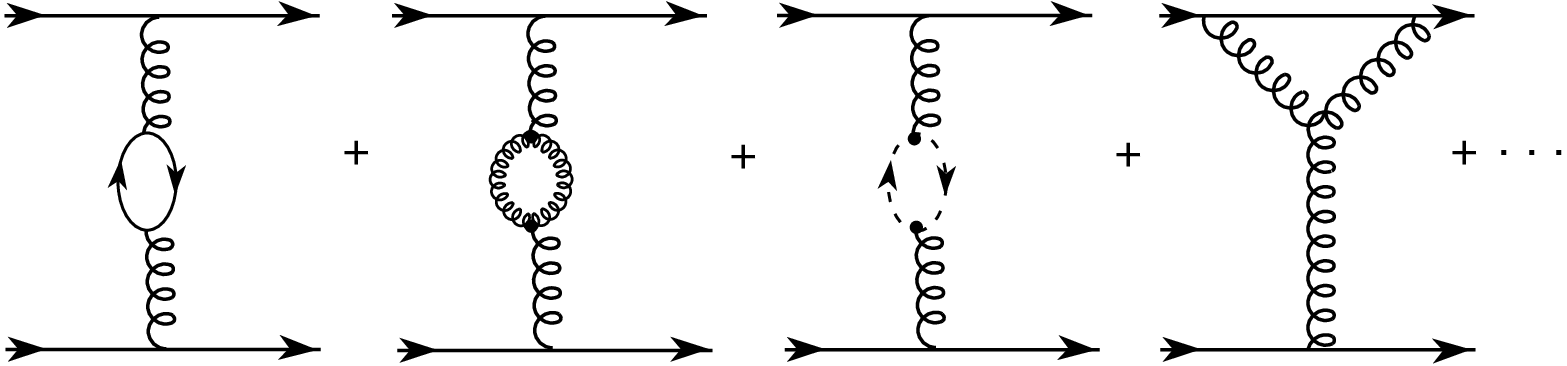}} 
\caption{Feynman diagrams contributing to the renormalization of the
strong coupling.}
%The dashed loop indicates the {\it ghost} correction discussed in Appendix B.}
\label{fig:g_si}
\end{figure}
%%%%%%%%%%%%%%%%%%%%%%%%%%%%%%%%%%%%%%%%%%%%%

The renormalization of the QCD coupling proceeds in a similar
way. Owing to the non-abelian character of $SU(3)_C$, there are additional
contributions involving gluon self-interactions. From the 
calculation of the relevant one-loop diagrams,
one gets the value of the first $\beta$ function 
coefficient \cite{GW:73}:
\bel{eq:QCD_beta}
\beta_1 \, = \, {2\over 3}\, T_F N_f  -  {11\over 6}\, C_A
\, = \, {2\, N_f - 11\, N_C \over 6} \, .
\ee
The positive contribution proportional to 
the number of quark flavours $N_f$ is generated by the
$q$--$\bar q$ loops and corresponds to the QED result 
(except for the $T_F=\frac{1}{2}$ factor).
The gluonic self-interactions introduce the additional {\it negative}
contribution proportional to $N_C$. This second term is responsible for the
completely different behaviour of QCD:\ $\beta_1 < 0$ if $N_f \leq 16$.
The corresponding QCD running coupling, $\alpha_s(Q^2)$,
decreases at short distances:    % \ie
\bel{eq:af_limit}
\lim_{Q^2\to\infty} \, \alpha_s(Q^2) \, = \, 0 \, .
\ee
Thus, for $N_f\leq 16$, QCD has indeed the required property of asymptotic 
freedom.
The gauge self-interactions of the gluons {\it spread out} the QCD charge,
generating an {\it antiscreening} effect. This could not happen in QED, because
photons do not carry electric charge. Only non-abelian gauge theories,
where the intermediate gauge bosons are self-interacting particles, have this
antiscreening property \cite{CG:73}.

Although quantum effects have introduced a dependence with the energy, we
still need a reference scale to decide when a given $Q^2$ can be
considered large or small. An obvious possibility is to choose the scale
at which $\alpha_s$ enters into a strong coupling regime (\ie\
$\alpha_s\sim 1$), where perturbation theory is no longer valid.
A more precise definition can be obtained from the solution of the
$\beta$-function differential equation \eqn{eq:beta}. At one loop, one gets
\bel{eq:Lambda_def}
\ln{\mu} + {\pi\over\beta_1\alpha_s(\mu^2)} \, = \, 
\ln{\Lambda} \, ,
\ee
where $\ln{\Lambda}$ is just an integration constant. Thus,
\bel{eq:alpha_Lambda}
\alpha_s(\mu^2) \, = \, 
{2\pi\over -\beta_1 \ln{\left({\mu^2/\Lambda^2}\right)}} \, .
\ee
In this way, we have traded the dimensionless parameter $g_s$ by the
dimensionful scale $\Lambda$.
The number of QCD free parameters is the same (1 for massless quarks), but
quantum effects have generated an energy scale.
Although, Eq.~\eqn{eq:alpha_run} gives the impression that the 
scale dependence of
$\alpha_s(\mu^2)$ involves two parameters, $\mu_0^2$ and
$\alpha_s(\mu_0^2)$, only the combination \eqn{eq:Lambda_def}
is actually relevant.
% as explicitly shown in \eqn{eq:alpha_Lambda}.

When $\mu>>\Lambda$, $\alpha_s(\mu^2)\to 0$, so that we recover asymptotic
freedom. At lower energies the running coupling gets larger; for
$\mu\to\Lambda$, $\alpha_s(\mu^2)\to \infty$ and perturbation theory breaks
down. The scale $\Lambda$ indicates when  the strong coupling blows up.
Eq.~\eqn{eq:alpha_Lambda} suggests that confinement at low energies 
is quite plausible
in QCD; however, it does not provide a proof because
perturbation theory is no longer valid when $\mu\to\Lambda$.

\subsection{Higher Orders}

Higher orders in perturbation theory are much more important in QCD than 
in QED, because the coupling is much larger (at ordinary energies). 
The $\beta$ function is known to four loops; in the
$\overline{\mbox{\rm MS}}$ scheme, the computed higher-order coefficients 
take the values \cite{CA:74}: ($\zeta_3 = 1.202056903 \ldots$)
\beqn\label{eq:beta_hl}
\beta_2 &=& -{51\over 4} + {19\over 12} \, N_f \, ; \qquad\quad
\beta_3 = {1\over 64}\left[ -2857 + {5033\over 9} \, N_f
- {325\over 27} \, N_f^2 \right] \, ;
\CR
\beta_4 &=&  \frac{-1}{128} \left[
\left( \frac{149753}{6}+3564 \,\zeta_3 \right)
-\left( \frac{1078361}{162}+\frac{6508}{27}\,\zeta_3 \right) N_f
 \right. \no \\ & & \quad\;\:\left.\mbox{}
+ \left( \frac{50065}{162}+\frac{6472}{81}\,\zeta_3 \right) N_f^2
+\frac{1093}{729}\, N_f^3 \right]~.
\eeqn
If $N_f\leq 8$, $\beta_2 <0$ \ ($\beta_3 < 0$ for $N_f\leq 5$, while
$\beta_4$ is always negative)
which further reinforces the asymptotic freedom behaviour.

The scale dependence of the running coupling at higher orders is given by:
$$
\alpha_s(\mu^2) \, =\, \alpha_s(\mu_0^2) \,\left\{ 1 - 
{\beta_1\over 2}{\alpha_s(\mu_0^2)\over\pi} 
\ln{\left({\mu^2\over\mu_0^2}\right)}
-{\beta_2\over 2}\left({\alpha_s(\mu_0^2)\over\pi}\right)^2 
\ln{\left({\mu^2\over\mu_0^2}\right)} + \cdots \right\}^{-1} \!\! . 
$$

\setcounter{equation}{0}
\section{Perturbative QCD Phenomenology}
\label{sec:QCDpert}

\subsection{$e^+e^-\to\; $ Hadrons}
\label{subsec:Rhadrons}

%%%%%%%%%%%%%%% FIGURE %%%%%%%%%%%%%%%%%%%%%
\begin{figure}[tbh] 
\myfigure{\epsfxsize=10cm \epsfbox{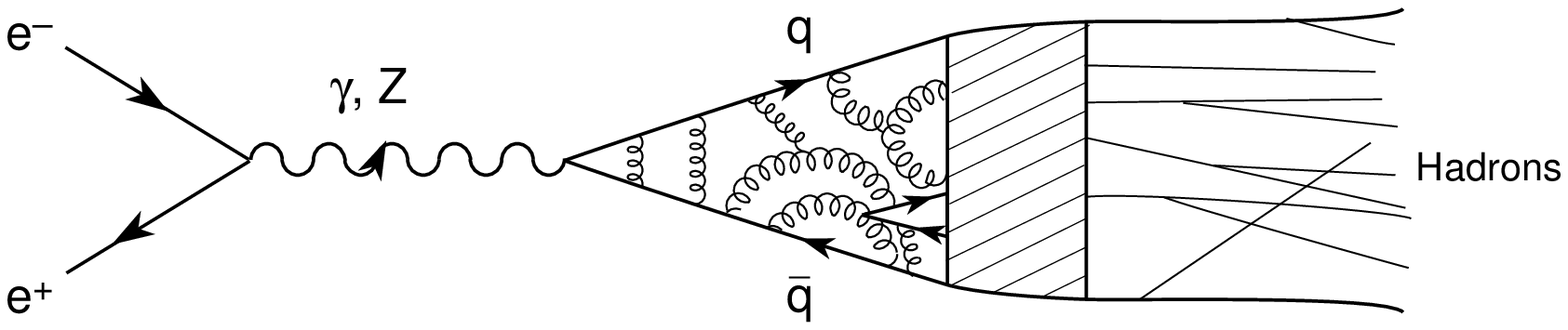}} 
\caption{$e^+e^-\to\gamma^*,Z^*\to\, $ hadrons.}
\label{fig:eeqqhad}
\end{figure}
%%%%%%%%%%%%%%%%%%%%%%%%%%%%%%%%%%%%%%%%%%%%%%

 The inclusive production of hadrons in $e^+e^-$ annihilation is a good
process for testing perturbative QCD predictions. 
The hadronic production occurs through the basic mechanism
$e^+e^-\to\gamma^*,Z^*\to q\bar q$,
where the final $q$-$\bar q$ pair interacts through the QCD forces;
thus, the quarks exchange and emit gluons (and $q'$-$\bar q'$ pairs) in all
possible ways.

At high energies, where $\alpha_s$ is small, we can use perturbative
techniques to predict the different subprocesses:
$e^+e^-\to q\bar q,q\bar q G, q\bar q GG, \ldots$\
 However, we do not
have a good understanding of the way quarks and gluons hadronize.
Qualitatively, quarks and gluons are created by the 
$q$-$\bar q$ current at very short distances, $x\sim 1/\sqrt{s}$. Afterwards,
they continue radiating additional soft gluons with smaller energies.
At larger
distances, $x\sim 1/\Lambda$, the interaction becomes strong and the
hadronization process occurs. Since we are lacking a rigorous description  of
confinement, we are unable to provide precise predictions of the
different exclusive processes, such as $e^+e^-\to 16\pi$. However, we can make
a quite accurate prediction for the total inclusive production of hadrons:
\bel{eq:sigma_total}
\sigma(e^+e^-\to\mbox{\rm hadrons}) = \sigma(e^+e^-\to
q\bar q + q\bar q G + q\bar q GG + \ldots) \, .
\ee
The details of the final hadronization are irrelevant for the inclusive sum,
because the probability to hadronize is just one owing to our confinement
assumption.

%%%%%%%%%%%%%%%  FIGURE %%%%%%%%%%%%%%%%%%%
\begin{figure}[tb] %[bht] %\vspace{4cm}
\myfigure{\hbox{\Large $\sigma\quad\;\sim\quad\; $}
\hbox{\epsfxsize=8.7cm \epsfbox[95 400 524 491]{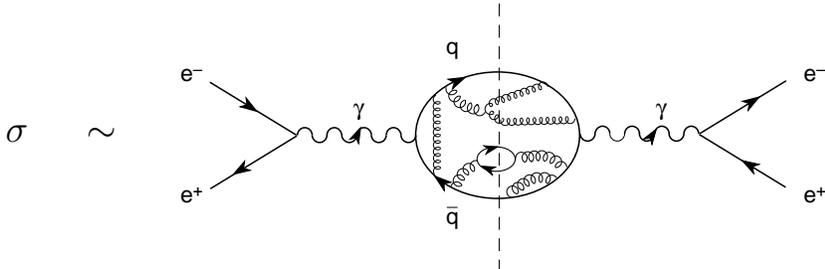}}}
\vspace{1.3cm}
\caption{Diagrammatic relation between the total hadronic production 
cross-section
and the two-point function $\Pi^{\mu\nu}(q)$.
The $q\bar q$ blob contains all possible QCD corrections. The dashed vertical 
line
indicates that the blob is cut in all possible ways, so that the left and
right sides correspond to the production amplitude $T$ and its 
complex conjugate
$T^\dagger$, respectively, for a given intermediate state.}
\label{fig:PiBlob}
\end{figure}
%%%%%%%%%%%%%%%

Well below the $Z$ peak, the hadronic production is dominated by the 
$\gamma$-exchange contribution. 
Thus, we can compute the cross-sections of all subprocesses
$e^+e^-\to\gamma^*\to q\bar q, q\bar q G,\ldots $
(at a given order in $\alpha_s$), and make the sum. Technically, 
it is much easier
to compute the QCD T-product of two electromagnetic currents
[$J^\mu_{\mbox{\rms em}} = \sum_f Q_f\, q_f\gamma^\mu q_f$]:
$$
\Pi^{\mu\nu}(q)\equiv i \int d^4x \; \e^{iqx}\;
\langle 0| T\left( J^\mu_{\mbox{\rms em}}(x) 
J^\nu_{\mbox{\rms em}}(0)^\dagger\right)|0\rangle
= \left(-g^{\mu\nu}q^2 + q^\mu q^\nu\right) \Pi_{\mbox{\rms em}}(q^2) \, .
$$
As shown in Fig.~\ref{fig:PiBlob}, the absorptive part of this object
(\ie\ the imaginary part, which results from cutting 
---putting {\it on shell\/}---
the propagators of the intermediate exchanged quarks and gluons in all
possible ways) just corresponds to the sum of the squared moduli of the 
different production amplitudes. The exact relation with the total
cross-section is:
\bel{eq:R_ee_QCD}
R_{e^+e^-} \equiv 
{\sigma(e^+e^-\to \mbox{\rm hadrons})\over\sigma(e^+e^-\to\mu^+\mu^-)} 
= 12 \pi \;\mbox{\rm Im}\Pi_{\mbox{\rms em}}(s) 
\, .
\ee

Neglecting the small (away from thresholds) quark mass
corrections, the ratio
$R_{e^+e^-}$ is given by a perturbative series in powers of $\alpha_s(s)$:
\beqn\label{eq:R_series}
\lefteqn{R_{e^+e^-} = \left(\sum_{f=1}^{N_f} Q_f^2\right)
\, N_C \, \left\{ 1 +
\sum_{n\geq 1} F_n \left({\alpha_s(s)\over\pi}\right)^{\! n} \right\} 
} &&\CR
&& = \left(\sum_{f=1}^{N_f} Q_f^2\right)
 \, N_C \, \Biggl\{ 1 +
F_1 {\alpha_s(\mu^2)\over\pi} +
\left[ F_2 + F_1 {\beta_1\over 2} \ln\left({s\over\mu^2}\right)\right]
\left({\alpha_s(\mu^2)\over\pi}\right)^{\! 2} 
\Biggr. 
\CR
& & \Biggl. \mbox{} +
\left[ F_3 + F_2 \beta_1 \ln\left({s\over\mu^2}\right) + F_1
\left( {\beta_2\over 2} \ln\left({s\over\mu^2}\right)
 + {\beta_1^2\over 4} \ln^2\left({s\over\mu^2}\right)\right)\right]
\left({\alpha_s(\mu^2)\over\pi}\right)^{\! 3} 
\Biggr. 
\CR
& & \Biggl. \mbox{} + 
\cO(\alpha_s^4) \Biggr\} .
\eeqn
The second expression, shows explicitly how the running coupling
$\alpha_s(s)$ sums an infinite number of
higher-order logarithmic terms.
So far, the calculation has been performed to order $\alpha_s^3$, with
the result (in the $\overline{\rm MS}$ scheme) \cite{GKL:91}:
\beqn\label{eq:F_results}
F_1 & = & 1 \, , \qquad\quad
F_2  =  1.986 - 0.115 \, N_f \, , \CR
F_3 & = & -6.637 - 1.200 \, N_f - 0.005 \, N_f^2 - 1.240\, 
{\left(\sum_f Q_f \right)^2\over 3 \sum_f Q_f^2 } \, . 
\eeqn
The different charge dependence on the last term is due to the
contribution from three intermediate gluons (with a separate quark trace
attached to each electromagnetic current in Fig.~\ref{fig:PiBlob}).
For $N_f=5$ flavours, one has:
$$
R_{e^+e^-}(s) \, = \, {11\over 3} \left\{ 1 + {\alpha_s(s)\over\pi} +
1.411 \left({\alpha_s(s)\over\pi}\right)^2
-12.80 \left({\alpha_s(s)\over\pi}\right)^3 + \cO(\alpha_s^4) \right\} \, .
$$

The perturbative uncertainty of this prediction is of order
$\alpha_s^4$ and
% since the coefficient $F_4$ is unknown. This uncertainty also
includes ambiguities related to the choice of renormalization scale and
scheme. Although, the total sum of the perturbative series is
independent of our renormalization conventions, different choices of scale
and/or scheme lead to slightly different numerical predictions for the
truncated series. For instance,
the perturbative series truncated at a finite order $N$
has an explicit scale dependence of order $\alpha_s^{N+1}$.
The numerical values of $\alpha_s$ and the $F_n$ ($n\geq 2$) coefficients 
depend on our choice of scheme
(also $\beta_n$ for $n > 2$).

The prediction for $R_{e^+e^-}(s)$ above the $b$-$\bar b$
threshold is compared with the data \cite{PDG:98}
in Fig.~\ref{fig:Ree},
taking into account mass corrections and $Z$--exchange
contributions. The two curves correspond to
$\Lambda^{(N_f=5)}_{\overline{\mbox{\rms MS}}} = 60$ MeV (lower curve)
and 250 MeV (upper curve). The rising at large energies is due to the
tail of the $Z$ peak.
A global fit to all data between 20 and 65 GeV yields \cite{HA:93}
\bel{eq:alpha_ee}
\alpha_s(35\:\mbox{\rm GeV}) \, = \, 0.146\pm0.030\, .
\ee

\subsection{$Z\to\; $ Hadrons}
\label{subsec:RZ}

The hadronic width of the $Z$ boson can be analyzed in a similar way:
$$
R_Z\equiv{\Gamma(Z\to\mbox{\rm hadrons})\over \Gamma(Z\to e^+e^-)} =
R_Z^{\mbox{\rms EW}} N_C \left\{ 1 +
\sum_{n\geq 1} \tilde{F}_n \left({\alpha_s(M_Z^2)\over\pi}\right)^n 
+ \cO\left({m_f^2\over M_Z^2}\right) \right\} \, .
$$
The global factor
\bel{eq:R_Z_ew}
R_Z^{\mbox{\rms EW}}\, =\, {\sum_f (v_f^2 + a_f^2)\over v_e^2 + a_e^2}\;
(1 + \delta_{\mbox{\rms EW}} )
\ee
contains the underlying electroweak $Z\to \sum_f q_f \bar q_f$ 
decay amplitude.
Since both vector and axial-vector couplings are present, the QCD correction
coefficients $\tilde{F}_n$ are slightly different from  $F_n$ for $n\geq 2$.
%For instance, the $Z$ axial coupling generates the two-loop contribution
%$Z\to t\bar t\to GG \to q\bar q$ (through triangular quark diagrams),
%which is absent in the vector case.
%this leads to an additional
%$\cO\left(\alpha_s^2 m_t^2/M_Z^2\right)$ correction.

To determine $\alpha_s$ from $R_Z$, one performs a global
analysis of the LEP/SLC data, taking properly into account
the higher-order electroweak corrections.
The latest $\alpha_s$ value reported by the LEP Electroweak Working Group
\cite{LEPEWG:99} is
\bel{eq:alpha_Z_fit}
\alpha_s(M_Z^2) \, = \, 0.119 \pm 0.003 \, .
\ee

\subsection{$\tau^-\to\nu_\tau + $ Hadrons}
\label{subsec:TauHadrons}

The calculation of QCD corrections to the inclusive hadronic
decay of the $\tau$ lepton \cite{BNP:92}
looks quite similar from a diagrammatic point of view. One puts
all possible gluon (and $q\bar q$) corrections to the basic decay diagram
in Fig.~\ref{fig:tau}, and computes the sum of all relevant partonic
subprocesses.
As in the $e^+e^-$ case, the calculation is more efficiently performed 
through the two-point function
\beqn\label{eq:PiTau}
\Pi_L^{\mu\nu}(q)& \equiv & i \int d^4x \, \e^{iqx}\;
\langle 0| T\left( L^\mu(x) L^\nu(0)^\dagger\right)|0\rangle
\CR 
& = & \left(-g^{\mu\nu}q^2 + q^\mu q^\nu\right) \Pi_L^{(1)}(q^2) 
\, + \, q^\mu q^\nu \,\Pi_L^{(0)}(q^2)\, ,
\eeqn
which involves the T-product of two left-handed currents,
$L^\mu = \bar u \gamma^\mu (1-\gamma^5) d_\theta$.
This object can be easily visualized through a diagram analogous to
Fig.~\ref{fig:PiBlob}, where the photon is replaced by a $W^-$ line and
one has a $\tau\nu_\tau$ pair in the external fermionic lines instead of
the $e^+e^-$ pair. 
The precise relation with the ratio $R_\tau$, defined in \eqn{eq:R_tau},
is:
$$ 
R_\tau  = 12 \pi \int^{m_\tau^2}_0 {ds \over m_\tau^2 } \,
 \left(1-{s \over m_\tau^2}\right)^2 
\left\{ \left(1 + 2 {s \over m_\tau^2}\right) 
 \mbox{\rm Im} \,\Pi^{(1)}_L(s)
 + \mbox{\rm Im} \,\Pi^{(0)}_L(s) \right\} \,  . 
$$

The three-body character of the basic decay mechanism, 
$\tau^-\to\nu_\tau u d_\theta$, shows here a crucial difference with
$e^+e^-$ annihilation. One needs to integrate over all possible neutrino
energies or, equivalently, over all possible values of the total hadronic
invariant--mass $s$. The spectral functions
$\mbox{\rm Im} \,\Pi^{(0,1)}_L(s)$ contain the dynamical information on the
invariant--mass distribution of the final hadrons.
The lower integration limit corresponds to the threshold for hadronic
production, \ie\ $m_\pi$ (equal to zero for massless quarks). Clearly, this
lies deep into the non-perturbative region where confinement is crucial.
Thus, it is very difficult to make a reliable prediction for
the $s$ distribution.

%%%%%%%%%%%%%%% FIGURE %%%%%%%%%%%%%%%%%%%%%%
\begin{figure}[tb]
\myfigure{\epsfysize =5cm \epsfbox{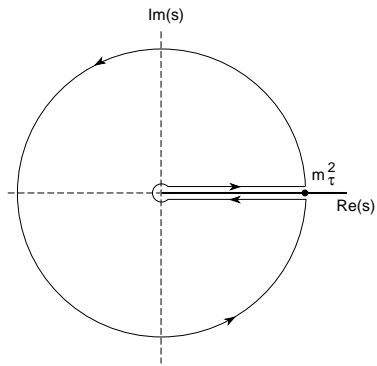}}
\vspace{-0.5cm}
\caption{Integration contour in the complex $s$-plane used to
compute $R_\tau$.}
% obtain Eq.~\protect\eqn{eq:circle}.}
\label{fig:circle}
\end{figure}
%%%%%%%%%%%%%%%%%%%%%%%%%%%%%%%%%%%%%%%%%%%%%%%%%

Fortunately, we have precious  non-perturbative information on
the dynamical functions $\Pi^{(0,1)}_L(s)$,
which allows us to accurately predict $R_\tau$. $\Pi^{(0,1)}_L(s)$ 
are analytic functions in the complex $s$-plane except for a
cut in the positive real axis. The physics we are interested in lies 
in the singular region, where hadrons are produced. 
We need to know the integral along the physical cut of 
$\,\mbox{\rm Im} \Pi^{(0,1)}_L(s) = -{i\over 2}
[\Pi^{(0,1)}_L(s+i\epsilon)-\Pi^{(0,1)}_L(s-i\epsilon)]$.
However, we can use Cauchy's theorem (close integrals of analytic functions are
zero if there are no singularities within the integration contour)
to express \cite{BNP:92}
$R_\tau$ as a contour integral in the complex $s$-plane running
counter-clockwise around the circle $|s|=m_\tau^2$:
$$
 R_\tau =
6 \pi i \oint_{|s|=m_\tau^2} {ds \over m_\tau^2} \,
 \left(1 - {s \over m_\tau^2}\right)^2
 \left\{ \left(1 + 2 {s \over m_\tau^2}\right) \Pi^{(0+1)}_L(s)
         - 2 {s \over m_\tau^2} \Pi^{(0)}_L(s) \right\} . 
$$
The advantage of this expression 
is that it requires dynamical information only for
complex $s$ of order $m_\tau^2$, which is significantly larger than the scale
associated with non-perturbative effects in QCD.  A perturbative
calculation of $R_\tau$ is then possible.

Using the so-called {\it Operator Product Expansion} techniques            
it is possible to show \cite{BNP:92}
that non-perturbative contributions are
very suppressed [$\sim (\Lambda/m_\tau)^6$].
Thus, $R_\tau$ is a perfect observable for determining the strong coupling.
 In fact, $\tau$ decay is probably the lowest energy process from which the
 running coupling constant can be extracted cleanly, without hopeless
 complications from non-perturbative effects. 
The $\tau$ mass lies fortuitously
 in a {\it compromise} region  where the coupling constant
 $\alpha_s$ is large enough that $R_\tau$ is very sensitive to its
 value, yet still small enough that the perturbative expansion
 still converges well. 

The explicit calculation gives \cite{BNP:92}:
\bel{eq:r_tau_total}
R_{\tau}  = 
 3 \left( |V_{ud}|^2 + |V_{us}|^2 \right)
S_{\mbox{\rms EW}} \left\{ 1 + \delta_{\mbox{\rms EW}}' + 
\delta_{\mbox{\rms P}} + \delta_{\mbox{\rms NP}}\right\}
\, , 
\ee
where  $S_{\mbox{\rms EW}} = 1.0194$ and 
$\delta_{\mbox{\rms EW}}' = 0.0010$
are the leading and next-to-leading electroweak corrections
\cite{tauEW}, and
$\delta_{\mbox{\rms P}}$ contains the dominant perturbative QCD contribution:
\beqn\label{eq:delta_0}
\delta_{\mbox{\rms P}} & =& {\alpha_s(m_\tau^2)\over\pi} +
\left[ F_2 - {19\over 24}\beta_1\right] 
\left( {\alpha_s(m_\tau^2)\over\pi} \right)^2 
\no\\ &  & \mbox{} +
\left[ F_3' - {19\over 12} F_2\beta_1 - {19\over 24}\beta_2 +
{265\over 288} \beta_1^2\right]
\left( {\alpha_s(m_\tau^2)\over\pi} \right)^3 + \cO(\alpha_s^4) \\
 & = & {\alpha_s(m_\tau^2)\over\pi} +
5.2023 \left( {\alpha_s(m_\tau^2)\over\pi} \right)^2
+ 26.366 \left( {\alpha_s(m_\tau^2)\over\pi} \right)^3
+ \cO(\alpha_s^4) \, .
\no\eeqn
The remaining factor $\delta_{\mbox{\rms NP}}$  %\approx -0.003\pm 0.003$
includes the small mass corrections and non-perturbative contributions.

Owing to its high sensitivity \cite{BNP:92} to $\alpha_s$
the ratio $R_\tau$ has been a subject of intensive study.
Many different sources of possible perturbative and non-perturbative
contributions have been analyzed in detail.
Higher-order logarithmic corrections have been resummed \cite{LDP:92a},
leading to very small renormalization--scheme dependences. 
The size of the non-perturbative contributions has been
experimentally analyzed, through a study of the invariant--mass
distribution of the final hadrons \cite{LDP:92b}; the data
imply \cite{tauNP} that
$\delta_{\mbox{\rms NP}}$ is smaller than 1\%.

The present experimental measurements \cite{tauNP} give
$\delta_{\mbox{\rms P}} = 0.200\pm 0.013$, which corresponds
(in the $\overline{\mbox{\rm MS}}$ scheme) to \cite{lp99tau}
\bel{alpha_s_tau}
\alpha_s(m_\tau^2)\, = \, 0.345\pm 0.020 \, ,
\ee
significantly larger ($11 \sigma$) than the
$Z$ decay measurement \eqn{eq:alpha_Z_fit}.
After evolution up to the scale $M_Z$, the strong coupling constant in
Eq.~\eqn{alpha_s_tau} decreases to
$\alpha_s(M_Z^2) = 0.1208\pm 0.0025$, in excellent agreement
with the direct measurement at the $Z$ peak and with a similar error bar.
The comparison of these two determinations of $\alpha_s$ in two
extreme energy regimes, $m_\tau$ and $M_Z$,
provides a beautiful test of the predicted running of
the QCD coupling; \ie\ a very significant experimental verification
of {\it asymptotic freedom}.

\subsection{$e^+e^-\to\; $ Jets}
\label{subsec:jets}

%%%%%%%%%%%%%%%  FIGURE  %%%%%%%%%%%%%%%%%%
\begin{figure}[tbh] 
\myfigure{\epsfxsize = 10cm \epsfbox{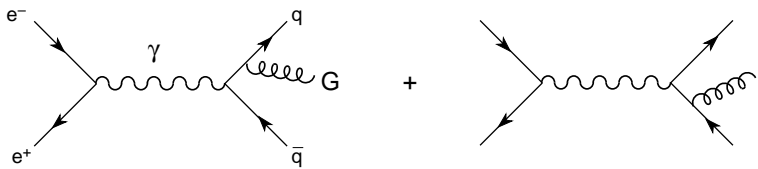}}
\caption{Gluon bremsstrahlung corrections to $e^+e^-\to q\bar q$.}
\label{fig:ee3J}
\end{figure}
%%%%%%%%%%%%%%%%%%%%%%%%%%%%%%%%%%%%%%%%%

At lowest order in $\alpha_s$,    %   the strong coupling, 
the hadronic production in
$e^+e^-$ collisions proceeds through $e^+e^-\to q\bar q$. Thus,
at high energies, the final hadronic states are predicted to 
have mainly a two-jet structure, which agrees with the
empirical observations.
At $\cO(\alpha_s)$, the emission of a hard gluon from a quark leg
%(see Fig.~\ref{fig:ee3J})
generates the $e^+e^-\to q\bar q G$ transition, leading to
3-jet configurations. For massless quarks, the differential
distribution of the 3-body final state is given by:
\bel{eq:3J_dis}
{1\over\sigma_0}\, {d^2\sigma\over dx_1 dx_2} \, = \,
{2\alpha_s\over 3\pi}\, {x_1^2 + x_2^2 \over (1-x_1) (1-x_2)} \ ,
\ee
where
\bel{eq:sigma0}
\sigma_0 \,\equiv\, {4\pi\alpha^2\over 3 s} \, N_C \,
\sum_{f=1}^{N_f} Q_f^2
\ee
is the lowest-order $e^+e^-\to\gamma^*\to q\bar q$ \ cross-section.
The kinematics is defined through the invariants
$s\equiv q^2$ and
$s_{ij}\equiv (p_i + p_j)^2 = (q-p_k)^2 \equiv s (1-x_k)$
($i,j,k = 1,2,3$), where $p_1$, $p_2$ and $p_3$ are the quark, 
antiquark and gluon momenta, respectively,
and $q$ is the total $e^+e^-$ momentum.
For given $s$, there are only
two independent kinematical variables since
\bel{eq:x_rel}
x_1 + x_2 + x_3 = 2 \, .
\ee
In the centre-of-mass system,  %[$q^\mu = (\sqrt{s}, \vec{0}\, )$],
$x_i = E_i/E_e = 2 E_i/\sqrt{s}$.

Eq.~\eqn{eq:3J_dis} diverges as $x_1$ or $x_2$ tend to 1.
This is a very different infinity from the ultraviolet ones
encountered before in the loop integrals.
In the present case, the tree amplitude itself is becoming singular in
the phase--space boundary. The problem originates in the infrared
behaviour of the intermediate quark propagators:
\bel{eq:infrared}
\ba
x_1\to 1 \qquad\Longleftrightarrow\qquad 
(p_2+p_3)^2 = 2 \,(p_2\cdot p_3) \to 0 \ ;
\\
x_2\to 1 \qquad\Longleftrightarrow\qquad 
(p_1+p_3)^2 = 2\, (p_1\cdot p_3) \to 0 \ .
\ea\ee
There are two distinct kinematical configurations leading to
{\it infrared divergences}:
\begin{enumerate}
\item {\bf Collinear gluon}: The 4-momentum of the gluon is
parallel to that of either the quark or the antiquark.
This is also called a {\it mass singularity}, since the
divergence would be absent if either the gluon or the
quark had a mass
($p_3\| p_2$ implies $s_{23}=0$ if $p_2^2=p_3^2=0$).
\item {\bf Soft gluon}: $p_3\to 0$.
\end{enumerate}
In either case, the observed final hadrons
will be detected as a 2-jet configuration, because the $qG$ or
$\bar qG$ system cannot be resolved.
Owing to the finite resolution of any detector, 
it is not possible (not even in principle) to separate those
2-jet events generated by the basic $e^+e^-\to q\bar q$ process,
from $e^+e^-\to q\bar q G$ events with a collinear or soft gluon.
In order to resolve a 3-jet event, the gluon should have an energy
and opening angle (with respect to the quark or antiquark)
larger than the detector resolution.
The observable 3-jet cross-section will never include the
problematic region $x_{1,2}\to 1$; thus, it will be finite,
although its value will depend on the detector resolution and/or
the precise definition of {\it jet}
(\ie\ $\sigma$ depends on the chosen integration
limits).

On the other side, the 2-jet configurations will include both
$e^+e^-\to q\bar q\; $ and $e^+e^-\to q\bar q G$ with an unobserved
gluon. The important question is then the infrared behaviour of the
sum of both amplitudes.
The exchange of virtual gluons among the quarks
generate an $\cO(\alpha_s)$ correction to the $e^+e^-\to q\bar q$
amplitude:
\bel{eq:T_eeqq}
T[e^+e^-\to q\bar q] = T_0 + T_1 + \cdots
\ee
where $T_0$ is the lowest-order (tree-level) contribution,
$T_1$ the $\cO(\alpha_s)$ correction, and so on.
The interference of $T_0$ and $T_1$ gives rise to an
$\cO(\alpha_s)$ contribution to the $e^+e^-\to q\bar q$
cross-section.

%%%%%%%%%%%%%%%  FIGURE  %%%%%%%%%%%%%%%%%%
\begin{figure}[tbh] 
\myfigure{\epsfxsize = 10cm \epsfbox{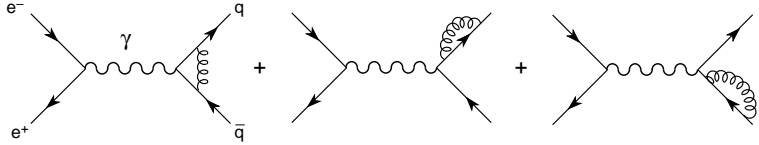}}
\caption{1-loop gluonic corrections to $e^+e^-\to q\bar q$.}
\label{fig:eeqq(g)}
\end{figure}
%%%%%%%%%%%%%%%%%%%%%%%%%%%%%%%%%%%%%%%%%

We know already that loop diagrams have ultraviolet divergences
which must be renormalized. In addition, they also have infrared
divergences associated with collinear and soft configurations
of the virtual gluon.
One can explicitly check that the $\cO(\alpha_s)$ infrared divergence
of $\sigma(e^+e^-\to q\bar q)$ exactly cancels the one in
$\sigma(e^+e^-\to q\bar q G)$, so that the sum is well-defined:
\bel{eq:infrared_sum}
\sigma(e^+e^-\to q\bar q) + \sigma(e^+e^-\to q\bar q G) + \cdots
= \sigma_0\,\left( 1 + {\alpha_s\over\pi} + \cdots\right) \ .
\ee
This is precisely the inclusive result discussed
in Sect.~\ref{subsec:Rhadrons}.
This remarkable cancellation of infrared divergences is
actually a general result (Bloch--Nordsieck \cite{BN:37}
and Kinoshita--Lee--Nauenberg \cite{KLN:62} theorems):
for inclusive enough cross-sections both the
soft and collinear infrared divergences cancel.

%%%%%%%%%%%%%%%
\begin{figure}[tbh] 
\myfigure{\epsfysize = 3cm \epsfbox{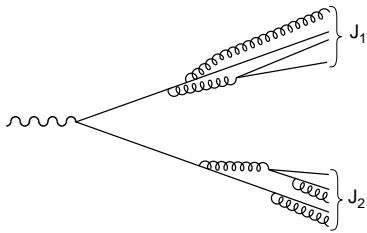}} 
\caption{2-jet configuration.}
\label{fig:jetdef}
\end{figure}
%%%%%%%%%%%%%%%

While the total hadronic cross-section is unambiguously defined,
we need a precise definition of jet in order to classify a given
event as a 2-, 3-, \ldots, or n-jet configuration.
Such a definition should be free of infrared singularities, and
insensitive to the details of the non-perturbative fragmentation
into hadrons.
A popular example of jet definition is the so-called
JADE algorithm \cite{JADE:86}, which makes use of an invariant--mass 
cut $y\/$:
\bel{eq:jade}
\mbox{\rm 3 jet} \qquad\Longleftrightarrow\qquad
s_{ij} \equiv (p_i + p_j)^2 > y s \qquad (\forall i,j=1,2,3) \ .
\ee

Clearly, both the theoretical predictions and the experimental
measurements depend on the adopted jet definition.
With the JADE algorithm, the fraction of 3-jet events
is predicted to be:
\beqn\label{eq:R3}
R_3  &=&  {2\alpha_s\over 3\pi} \,\biggl\{
(3-6y) \,\ln{\!\left({y\over 1-2y}\right)}
+ 2 \,\ln^2{\!\left({y\over 1-y}\right)} + {5\over 2}
-6\, y -\frac{9}{2}\, y^2
\biggr.\no\\ &&\biggl.\qquad \mbox{}
+ 4 \mbox{\rm Li}_2\!\left({y\over 1-y}\right) -{\pi^2\over 3}
\biggr\}  ,
\eeqn
where
\bel{eq:Li}
\mbox{\rm Li}_2(z)\,\equiv\, -\int_0^z\,{d\xi\over 1-\xi}\,
\ln{\xi} \ .
\ee
The corresponding fraction of 2-jet events is given by
$R_2 = 1 - R_3$.
The fraction of 2- or 3-jet events obviously depends on the
chosen cut $y$. 
The infrared singularities are manifest in the divergent behaviour
of $R_3$ for $y\to 0$.

At higher orders in $\alpha_s$ one needs to define the
different multi-jet fractions. For instance, one can
easily generalize the JADE algorithm an classify a
$\{p_1,p_2,\ldots,p_n\}$ event as a n-jet configuration
provided that $s_{ij}>y s$ for all $i,j=1,\ldots,n$.
If a pair of momenta does not satisfy this constraint, they are
combined into a single momentum and the event is considered
as a $(n-1)$ jet configuration (if 
the constraint is satisfied by all other combinations
of momenta).
The general expression for the fraction of n-jet events takes the
form:
\bel{eq:R_n_frac}
R_n(s,y) \, = \, \left({\alpha_s(s)\over\pi}\right)^{n-2}
\sum_{j=0} C_j^{(n)}(y)\,\left({\alpha_s(s)\over\pi}\right)^j \ ,
\ee
with $\sum_n R_n = 1$.
A few remarks are in order here:
\bi
\item The jet fractions have a
high sensitivity to $\alpha_s$ [$R_n\sim \alpha_s^{n-2}$].
 Although the sensitivity 
increases with $n$, the number of events decreases
with the jet multiplicity.
\item Higher-order $\alpha_s(\mu^2)^j\ln^k(s/\mu^2)$ terms have
been summed into $\alpha_s(s)$. However, the coefficients
$C_j^{(n)}(y)$ still contain $\ln^k(y)$ terms.
At low values of $y$, the infrared divergence ($y\to 0$) reappears
and the perturbative series becomes unreliable. 
For large $y$, the jet fractions $R_n$ with $n\geq 3$ are small.
\item Experiments measure hadrons rather than partons. Therefore,
since these observables are not fully inclusive, there is an
unavoidable dependence on the non-perturbative fragmentation into
hadrons. This is usually modelled through Monte Carlo analyses,
and introduces theoretical uncertainties which need to be estimated.
\ei

Different jet algorithms and jet variables
(jet rates, event shapes, energy correlations, \ldots) have
been introduced to optimize the perturbative analysis.
In some cases, a resummation of
$\alpha_s(s)^n\ln^m\! (y)$ contributions with $m>n$ has been performed
to improve the predictions at low $y$ values \cite{CA:91}.

Many measurements of $\alpha_s$, using different jet variables, have
been performed. All results are in good agreement, providing a
consistency test of the QCD predictions.
An average of measurements at the $Z$ peak, from LEP and SLC,
using resummed $\cO(\alpha_s^2)$ fits to a large set of shape variables,
gives \cite{PDG:98}:
\bel{eq:alpha_s_shapes}
\alpha_s(M_Z^2)\, = \, 0.122\pm 0.007 \, . 
\ee
A recent $\cO(\alpha_s^2)$ DELPHI analysis \cite{DELPHIalpha}
of 18 different 
event--shape observables finds an excellent fit to the data,
by allowing the renormalization scale to be determined from the
fit; this procedure gives a more precise value:
$\alpha_s(M_Z^2)\, = \, 0.117\pm 0.003$.

Three-jet events can also be used to test the gluon spin.
For a spin-0 gluon, the differential distribution is still
given by Eq.~\eqn{eq:3J_dis}, but changing the $x_1^2+x_2^2$ factor
in the numerator to $x_3^2/4$.
In general, one cannot readily be sure which hadronic jet emerges
via fragmentation from a quark (or antiquark), and which from a gluon.
Therefore, one adopts instead a jet ordering, 
$x_1>x_2>x_3$, where $x_1$ refers to the most energetic jet, 2 to the next
and 3 to the least energetic one, which most likely would correspond to
the gluon.
When $x_2\to 1$ ($x_1\to 1$) the vector-gluon distribution is singular,
but the corresponding scalar-gluon distribution is not
because at that point
$x_3 = (1-x_1) + (1-x_2) \to 0$.
The measured distribution agrees very well with the QCD predictions with
a spin-1 gluon; a scalar gluon is clearly excluded.

The predictions for jet distributions and event shapes are
functions of the colour group factors $T_F=1/2$,
$C_F = (N_C^2-1)/(2 N_C)$ and $C_A=N_C$. These quantities,
defined in Eq.~\eqn{eq:invariants}, result from the
colour algebra associated with the different interaction vertices,
and characterize the colour symmetry group.
If the strong interactions were based on a different gauge group,
the resulting predictions would differ in the values of these
three factors.
Since the vertices contribute in a different way to
different observables, these colour factors can be measured by
performing a combined fit.
The data are in excellent agreement with the $SU(3)$ values,
and rule out the Abelian model and many classical
Lie groups.

\setcounter{equation}{0}
\section{Determinations of the Strong Coupling}

%%%%%%%%%%%%%%%%%%%%%  FIGURES %%%%%%%%%%%%%%%%%%%%%%%%%%
\begin{figure}[tbh]
\myfigure{\mbox{\epsfysize=8cm\epsffile{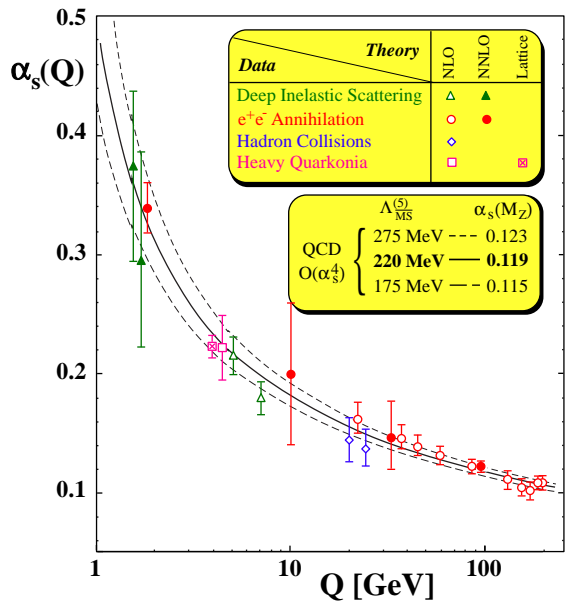}}}
\vspace{-0.3cm}
\caption{Compilation \protect\cite{BE:99} of $\alpha_s$ 
measurements as function of the energy scale.}
\label{fig:alpha_run}
\end{figure}
%%%%%%%%%%%%%%%%%%%%% End figures %%%%%%%%%%%%%%%%%%%%%%%%%%

%%%%%%%%%%%%%%%  FIGURE %%%%%%%%%%%%%%%%%
\begin{figure}[tbh]
\myfigure{\rotate[l]{\epsfxsize = 9cm \epsfbox{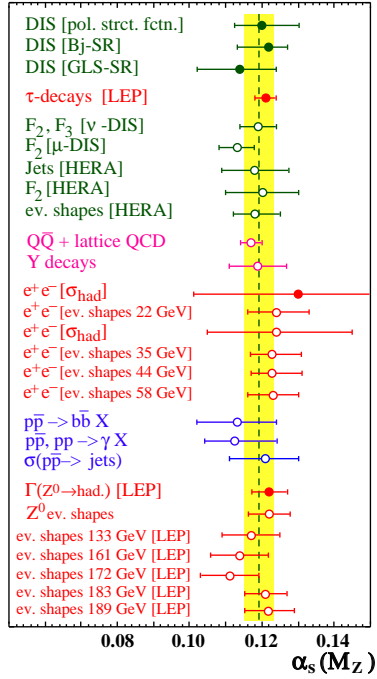}}}
\caption{Summary \protect\cite{BE:99}
of $\alpha_s$ measurements, evolved to the scale $M_Z$.}
\label{fig:alpha_Z}
\end{figure}
%%%%%%%%%%%%%%%%%%%%%%%%%%%%%%%%%%%%%%%%%

In the massless quark limit, QCD has only one free parameter: the
strong coupling $\alpha_s$. Thus, all strong interaction
phenomena should be described in terms of this single input.
The measurements of $\alpha_s$ at different processes and at
different mass scales provide then a crucial test of QCD.
Obviously, the test should be restricted to those processes
where perturbative techniques are reliable. Moreover, the same
definition of $\alpha_s$ should be taken
everywhere; the $\overline{\mbox{\rm MS}}$ scheme is usually adopted as
the standard convention.
Since the running coupling is a function of energy,
one can either compare the different determinations at the
different scales where they are measured, checking in this way
the predicted $Q^2$ dependence of the coupling, or
use this prediction to bring all measurements to a common
reference scale where they are compared. Nowadays, the $Z$ mass
scale is conventionally chosen for such a comparison.

In order to assess the significance of the test, it is very important
to have a good understanding of the uncertainties associated
with the different measurements. This is not an easy question,
because small non-perturbative effects can be present in many
observables. In addition, some quantities have been computed
to a very good perturbative accuracy (next-to-next-to-leading order), 
while others are only known at the leading or next-to-leading order;
the resulting values of $\alpha_s$ refer then to different
perturbative approximations.
The estimate of theoretical uncertainties is also affected by the
plausible asymptotic (\ie\ non-convergent) behaviour of the
perturbative series in powers of $\alpha_s$. Although this is a
common problem of Quantum Field Theories, it is probably more
severe in QCD because the coupling is rather large (at usual energies).

Fig.~\ref{fig:alpha_run} summarizes \cite{BE:99}
the most reliable measurements
of the strong coupling as function of the energy scale. The
agreement with the predicted running of $\alpha_s$,
indicated by the curves, is indeed very good.
Fig.~\ref{fig:alpha_Z} compares \cite{BE:99} the different 
measurements at the common reference scale $M_Z$. 
The average af all determinations gives \cite{PDG:98}:
\bel{eq:alpha_average}
\alpha_s(M_Z^2) \, = 0.119 \pm 0.002 \, .
\ee

\setcounter{equation}{0}
\section{Chiral Symmetry}
\label{sec:chpt}

Up to now, we have only discussed
those aspects of QCD which can be analyzed in a perturbative way.
Thus, we have restricted ourselves to the study of scattering processes at
large momentum transfers, and inclusive transitions
which avoid the hadronization problems.
The rich variety of strong--interacting phenomena governed by the confinement
regime of QCD has been completely ignored.

There are certainly many approximate tools to investigate
particular aspects of non-perturbative physics; 
however, rigorous first-principle QCD calculations seem
unfortunately out of reach for present techniques.
Nevertheless, we can still investigate some general properties of QCD
using symmetry considerations. 

\subsection{Flavour Symmetries}

In order to build the QCD Lagrangian,
we made extensive use of the $SU(3)_C$
colour symmetry, which is the basis of
the strong interaction dynamics. The Lagrangian \eqn{eq:L_QCD}
has additional {\it global} symmetries associated with the quark
flavour numbers:
\begin{enumerate}
\item $\cL_{\mbox{\rms QCD}}$ is invariant under a global phase
redefinition of all quark flavours,
\be\label{eq:baryon_number}
q_f \; \longrightarrow\; \exp(i\theta)\; q_f \ .
\ee
This symmetry is associated with the conservation of the baryon number.
\item $\cL_{\mbox{\rms QCD}}$ is also invariant
under independent phase redefinitions
of the different quark flavours,
\bel{eq:flavour_number}
q_f \; \longrightarrow\; \exp(i\theta_f)\; q_f \ .
\ee
This symmetry implies the conservation of flavour.
\item For equal quark masses, there is a larger symmetry
under $SU(N_f)$ transformations in flavour space,
\bel{eq:isospin}
q_f \;\longrightarrow\; U_{ff'}\; q_{f'} \ , \qquad\qquad U\in SU(N_f) \ .
\ee
This is a good symmetry of the light flavour sector ($u$, $d$, $s$), where
quark masses can be ignored in first approximation. One has then the
well-known isospin ($N_f=2$) and $SU(3)$ symmetries. 
\item
In the absence of quark masses, the QCD Lagrangian splits into
two independent quark sectors,
\bel{eq:LR_sectors}
\cL_{\mbox{\rms QCD}}
\,\equiv\, -{1\over 4}\, G^{\mu\nu}_aG_{\mu\nu}^a
+ i\,\bar q_L \gamma^\mu D_\mu q_L  +
i\,\bar q_R \gamma^\mu D_\mu q_R \ .
\ee
Here, $q$ denotes the flavour (and colour) vector \ 
$q^T = (u,d,\ldots)$, and $L,R$ stand for the
left- and right-handed components of the quarks.
Thus, the two quark chiralities live in separate
flavour spaces (gluon interactions
do not change the chirality), implying that all previous flavour
symmetries get duplicated in the two chiral sectors.
\end{enumerate}

The baryon number symmetry \eqn{eq:baryon_number} is usually called $U(1)_V$,
since both chiralities transform in the same way.
Its chiral replication is the corresponding $U(1)_A$ transformation:
\bel{eq:U1_A}
q_L \; \longrightarrow\; \exp(-i\theta)\; q_L \ ; \qquad\qquad
q_R \; \longrightarrow\; \exp(i\theta)\; q_R \ .
\ee
This symmetry of the classical (massless) QCD Lagrangian gets broken by
quantum effects (triangular loops of the type shown in Fig.~\ref{fig:triangle},
with gluons instead of photons); this is the so-called $U(1)_A$ anomaly.
Although \eqn{eq:U1_A} is not a true symmetry of QCD, 
it gets broken in a very
specific way, which leads to important implications. 
A discussion of the phenomenological role of anomalies is beyond the
scope of these lectures. However, let me mention that this anomaly is
deeply related to interesting low-energy phenomena such as the
understanding of the $\eta'$ mass, or
the polarized proton structure functions.
  
I want to concentrate here in the chiral extension of the old eightfold
$SU(3)_V$ symmetry, \ie\ in the global 
$G\equiv SU(3)_L\otimes SU(3)_R$ symmetry
of the QCD Lagrangian for massless $u$, $d$ and $s$ quarks.
This larger symmetry is not directly seen in the hadronic spectrum.
Although hadrons can be nicely classified in $SU(3)_V$ representations,
degenerate multiplets with opposite parity do not exist.
Moreover, the octet of pseudoscalar mesons 
($\pi$,$K$,$\eta$) happens to be much lighter than all other 
hadronic states.

There are two different ways in which a symmetry of the Lagrangian can
be realized. In the usual one (Wigner--Weyl), the ground state (the vacuum)
is also invariant. Then, all physical states can be classified in
irreducible representations of the symmetry group \cite{CO:66}.
Certainly, the hadronic spectrum does not look like that in the case
of the chiral group. 

There is a second (Nambu--Golstone), 
more sophisticated, way to realize a symmetry. In
some cases, the vacuum is not symmetric. The hadronic spectrum
corresponds to energy excitations over the physical vacuum and, therefore,
will not manifest the original symmetry of the Lagrangian.  
However, Goldstone's theorem \cite{GO:61} says that
in such a case there should appear a massless scalar for each
broken generator of the original symmetry group.
If the chiral symmetry is realized in this way, there should be
eight pseudoscalar massless states (Goldstone bosons) 
in the hadronic spectrum; this is precisely the number of states
of the lightest hadronic multiplet: the $0^-$ octet.
Thus, we can identify the $\pi$, $K$ and $\eta$ with the Goldstone
modes of QCD; their small masses being generated by the
quark mass matrix which explicitly breaks the global chiral symmetry
of the Lagrangian.

In the Standard electroweak model, the local 
$SU(2)_L\otimes U(1)_Y$
symmetry is also realized in
the Nambu--Goldstone way. There, the symmetry--breaking phenomena is
assumed to be related to the existence of some scalar multiplet which gets
a vacuum expectation value. Since a {\it local} symmetry gets 
(spontaneously) broken in that case, the Goldstone modes combine
with the gauge bosons giving massive spin-1 states plus the Higgs particle.
The QCD case is simpler, because it is a {\it global} symmetry the
one which gets broken. However, something should play the role of the
electroweak scalar field. Since quarks are the only fields carrying
flavour, they should be responsible for the symmetry breaking. The simplest
possibility is the appearance of a quark condensate
\bel{eq:quark_condensate}
v\equiv\langle 0| \bar u u |0\rangle =
\langle 0| \bar d d |0\rangle =
\langle 0| \bar s s |0\rangle < 0 \ ,
\ee
generated by the non-perturbative QCD dynamics. 
This would produce a dynamical breaking
of chiral symmetry, keeping at the same time the observed
$SU(3)_V$ symmetry.
 
\subsection{Effective Chiral Lagrangian}

The Goldstone nature of the pseudoscalar mesons implies strong
constraints on their interactions, which can be most easily analyzed
on the basis of an effective Lagrangian.
The Goldstone bosons correspond to the zero-energy excitations
over the quark condensate; their fields can be collected in a
$3\times 3$ unitary matrix $U(\phi)$,
\be
\langle 0| \bar q^j_L q^i_R|0\rangle \, \longrightarrow \, {v\over 2}\,
U^{ij}(\phi) ,
\ee
which parametrizes those excitations.
A convenient parametrization is given by
\be
U(\phi)  \equiv \exp{\left(i \sqrt{2} \Phi / f\right)} ,
\ee
where
\be
\ba
\Phi (x) \equiv
\frac{\dis \vec{\lambda}}{\dis \sqrt 2} \, \vec{\phi}
 = \, \left( \begin{array}{ccc}
\frac{\dis \pi^0}{\dis \sqrt 2} \, + \, \frac{\dis \eta_8}{\dis \sqrt 6}
 & \pi^+ & K^+ \\
\pi^- & - \frac{\dis \pi^0}{\dis \sqrt 2} \, + \, \frac{\dis \eta_8}
{\dis \sqrt 6}    & K^0 \\
K^- & \bar K^0 & - \frac{\dis 2 \, \eta_8}{\dis \sqrt 6}
\end{array}  \right) .
\ea
\ee
The matrix
$U(\phi)$ transforms linearly under the chiral group, [$g_{L,R}\in SU(3)_{L,R}$]
\be\label{eq:utransf}
q_L\,\stackrel{G}{\longrightarrow}\, g_L\, q_L ,\quad
q_R\,\stackrel{G}{\longrightarrow}\, g_R\, q_R \qquad
\Longrightarrow\qquad
U(\phi) \, \stackrel{G}{\longrightarrow}\, g_R \, U(\phi) \, g_L^\dagger \; ,
%\qquad [g_{L,R}\in SU(3)_{L,R}] ,
\ee
but the induced transformation on the Goldstone fields
$\vec{\phi}$ is highly non-linear.
 
Since there is a mass gap separating the pseudoscalar octet
from the rest of the hadronic spectrum,
we can build a low-energy  effective field theory 
containing only the Goldstone modes.
We should write the most general Lagrangian involving the matrix
$U(\phi)$, which is consistent with chiral symmetry.
Moreover, we can
organize the Lagrangian in terms of increasing powers of
momentum or, equivalently, in terms of an increasing number of
derivatives (parity conservation requires an even number of derivatives):
\be
{\cal L}_{\hbox{\rms eff}}(U) \, = \, \sum_n \, {\cal L}_{2n} \ .
\ee
In the low-energy domain, the
terms with a minimum number of derivatives will dominate.
 
Due to the unitarity of the $U$ matrix, $U U^\dagger = 1$, at least
two derivatives are required to generate a non-trivial interaction.
To lowest order, the effective chiral Lagrangian is uniquely
given by the term
\be\label{eq:l2}
{\cal L}_2 = {f^2\over 4}\,
\mbox{\rm Tr}\left[\partial_\mu U^\dagger \partial^\mu U \right] .
\ee

Expanding $U(\phi)$ in a power series in $\Phi$, one obtains the
Goldstone's kinetic terms plus a tower of interactions involving
an increasing number of pseudoscalars.
The requirement that the kinetic terms are properly normalized
fixes the global coefficient $f^2/4$ in (\ref{eq:l2}).
All interactions among the Goldstones can then be predicted in terms
of the single coupling $f$:
\be
{\cal L}_2 \, = \, {1\over 2} \,\mbox{\rm Tr}\left[\partial_\mu\Phi
\partial^\mu\Phi\right]
\, + \, {1\over 12 f^2} \,\mbox{\rm Tr}\left[
(\Phi\stackrel{\leftrightarrow}{\partial}_{\!\mu}\Phi) \,
(\Phi\stackrel{\leftrightarrow}{\partial^\mu}\Phi)
\right] \, + \, \cO(\Phi^6/f^4) .
\ee

To compute the $\pi\pi$ scattering amplitude, for instance, is now
a trivial perturbative exercise. One gets the well-known
Weinberg result \cite{WE:66}  
\be\label{eq:WE1}
T(\pi^+\pi^0\to\pi^+\pi^0) = {t/ f^2}
\qquad\qquad [t\equiv (p_+' - p_+)^2].
\ee
Similar results can be obtained for $\pi\pi\to 4\pi, 6\pi, 8\pi, \ldots$\,
The non-linearity of the effective Lagrangian relates
amplitudes with different numbers of Goldstone bosons, allowing
for absolute predictions in terms of $f$.
Notice that the Goldstone interactions are proportional to their
momenta (derivative couplings). 
Thus, in the zero-momentum limit, pions become free.
In spite of confinement, QCD has a weakly--interacting regime at low
energies, where a perturbative expansion in powers of momenta can be applied. 

It is straightforward to generalize the effective Lagrangian
(\ref{eq:l2}) to incorporate electromagnetic and semileptonic
weak interactions \cite{GL:85}. One learns then that $f$ is in fact
the pion decay constant $f\approx f_\pi=92.4$ MeV, measured in 
$\pi\to\mu\nu_\mu$ decay.
The corrections induced by the non-zero quark
masses are taken into account through the term
\bel{eq:L_m}
\cL_m \, = \, {|v|\over 2}\, \mbox{\rm Tr}\left[\cM (U + U^\dagger)\right]
\ , \qquad \cM\equiv\mbox{\rm diag}(m_u,m_d,m_s) \ ,
\ee
which breaks chiral symmetry in exactly the same way as
the quark mass term in the underlying QCD Lagrangian does.
Eq.~\eqn{eq:L_m}
gives rise to a quadratic pseudoscalar mass term plus
additional interactions proportional to the quark masses.
Expanding in powers of $\Phi$
(and dropping an irrelevant constant), one has
\be\label{eq:massterm}
\cL_m \, = \,
 |v|\,\left\{ -{1\over f^2} \mbox{\rm Tr}\left[{\cal M}\Phi^2\right]
+ {1\over 6 f^4} \mbox{\rm Tr}\left[ {\cal M} \Phi^4\right]
+ \cO(\Phi^6/f^6) \right\} .
\ee

The explicit evaluation of the trace in the quadratic mass term provides
the relation between the physical meson masses and the quark masses:
%
%\beqn\label{eq:masses}
$$
m_{\pi^\pm}^2  = (m_u+m_d) {|v|\over f^2}\ , 
\qquad\qquad
m_{\pi^0}^2  =  (m_u+m_d) {|v|\over f^2} - \varepsilon +
\cO(\varepsilon^2)\ , 
$$
\bel{eq:masses}
m_{K^\pm}^2  =  (m_u + m_s) {|v|\over f^2}\ , 
\qquad\qquad  
m_{K^0}^2  =  (m_d + m_s) {|v|\over f^2}\ , 
\ee
$$
m_{\eta_8}^2  = {1\over 3} (m_u+m_d + 4 m_s)  {|v|\over f^2} + \varepsilon +
\cO(\varepsilon^2)\ , 
$$
where
\be
\varepsilon = {|v|\over 2f^2}\ {(m_u - m_d)^2\over  (2m_s - m_u-m_d)} \ .
\ee
Chiral symmetry relates the magnitude of the meson and quark masses
to the size of the quark condensate.
Taking out the common $|v|/f^2$ factor, Eqs.~(\ref{eq:masses}) imply
the old Current Algebra mass ratios,
\be\label{eq:mratios}
{m^2_{\pi^\pm}\over m_u+m_d} = {m^2_{K^+}\over (m_u+m_s)} =
{m_{K^0}\over (m_d+m_s)}
\approx {3 m^2_{\eta_8}\over (m_u+m_d + 4 m_s)} \ ,
\ee
and
[up to $\cO(m_u-m_d)$ corrections]
the Gell-Mann--Okubo mass relation
\be
3 m^2_{\eta_8} = 4 m_K^2 - m_\pi^2 \ .
\ee

Although chiral symmetry alone cannot fix the absolute values
of the quark masses, it gives information about quark mass
ratios. Neglecting the tiny $\cO(\varepsilon)$ effects,
one gets the relations
\bel{eq:ratio1}
{m_d - m_u \over m_d + m_u} \, =\, 
{(m_{K^0}^2 - m_{K^+}^2) - (m_{\pi^0}^2 - m_{\pi^+}^2)\over m_{\pi^0}^2}
\, = \, 0.29 \ , 
\ee
\bel{eq:ratio2}
{2 m_s -m_u-m_d\over 2 (m_u+m_d)} \, = \, 
{m_{K^0}^2 - m_{\pi^0}^2\over m_{\pi^0}^2}
\, = \, 12.6 \ . 
\ee
In (\ref{eq:ratio1}) we have subtracted the pion square mass
difference, to take into account the electromagnetic contribution
to the pseudoscalar meson self-energies;
in the chiral limit ($m_u=m_d=m_s=0$), this contribution is proportional
to the square of the meson charge and it is the same for $K^+$ and $\pi^+$.
The mass formulae (\ref{eq:ratio1}) and (\ref{eq:ratio2})
imply the quark mass ratios advocated by Weinberg:
\be\label{eq:Weinbergratios}
m_u : m_d : m_s = 0.55 : 1 : 20.3 \ .
\ee
Quark mass corrections are therefore dominated by $m_s$, which is
large compared with $m_u$ and $m_d$.
Notice that the difference $m_d-m_u$ is not small compared with
the individual up  and down quark masses; in spite of that,
isospin turns out
to be an extremely good symmetry, because
isospin--breaking effects are governed by the small ratio
$(m_d-m_u)/m_s$.

The $\Phi^4$ interactions in (\ref{eq:massterm})
introduce mass corrections to the $\pi\pi$ scattering amplitude
(\ref{eq:WE1}),
\be\label{eq:WE2}
T(\pi^+\pi^0\to\pi^+\pi^0) = {\left(t - m_\pi^2\right)/ f_\pi^2} \ .
\ee
Since $f\approx f_\pi$ is fixed from pion decay, this result
is now an absolute prediction of chiral symmetry.
 
The lowest-order chiral Lagrangian encodes
in a very compact way all the Current Algebra results obtained in
the sixties \cite{currentalgebra}.
The nice feature of the chiral approach is its elegant
simplicity, which
allows to estimate higher-order corrections in a systematic way.
A detailed summary of the chiral techniques and their
phenomenological applications can be found in
Refs.~42.  %\cite{ChPT}. 

\setcounter{equation}{0}
\section{Quark Masses}
\label{sec:mq}

Owing to confinement, quarks are not asymptotic states of
QCD and, therefore, their masses cannot be directly measured.
They can only be determined indirectly, through their influence on
hadron properties.
Moreover, since they are not {\it observable quantities}, 
quark masses need to be properly defined; \ie\ their values
depend on the chosen conventions.

A possible (intuitive) definition is the so-called {\it pole mass\/}:
the pole of the perturbative quark propagator \cite{TA:81}.
However, there is no pole beyond perturbation theory. Therefore,
one expects this quantity to be very sensitive to non-perturbative
long-distance effects.

The simplest prescription is to consider $m_q$ in the same way as
$\alpha_s$: quark masses are just additional couplings in the
QCD Lagrangian \eqn{eq:L_QCD}. These couplings need to be
renormalized and, therefore, one gets scale--dependent
{\it running} quark masses. We will adopt the usual
$\overline{\mbox{\rm MS}}$ scheme to define $m_q(\mu^2)$.

The scale dependence of $m_q(\mu^2)$ is regulated by 
the so-called $\gamma$ function,
\bel{eq:gamma}
\mu\, {d m_q\over d\mu} \,\equiv\, - m_q \,\gamma(\alpha_s) \, ;
\qquad\qquad \gamma(\alpha_s)\, =\, \gamma_1 {\alpha_s\over \pi} +
\gamma_2 \left({\alpha_s\over\pi}\right)^2 + \cdots
\ee
which is known to four loops \cite{TA:81,gamma}:
\beqn
\gamma_1 &=& 2~, \qquad
\gamma_2 = \left[\frac{101}{12}-\frac{5}{18}\, N_f\right]~, 
\CR
\gamma_3 &=& \frac{1}{32} \left[1249+
\left( -\frac{2216}{27}-\frac{160}{3} \,\zeta_3 \right) N_f
-\frac{140}{81}\, N_f^2 \right]~, 
\CR
\gamma_4 &=& \frac{1}{128} \left[\frac{4603055}{162}
+\frac{135680}{27}\,\zeta_3 - 8800\,\zeta_5 \right. 
\CR 
&&\mbox{} + \left(-\frac{91723}{27}-\frac{34192}{9}\,\zeta_3 + 880\,\zeta_4
+\frac{18400}{9}\,\zeta_5 \right) N_f 
\CR 
&&\mbox{} \left. + \left( \frac{5242}{243}+\frac{800}{9}\,\zeta_3
-\frac{160}{3}\,\zeta_4\right) N_f^2
+\left( -\frac{332}{243}+\frac{64}{27}\,\zeta_3 \right) N_f^3 \right]~. 
\eeqn
Here $\zeta_n$ is the Riemann zeta function (
$\zeta_2=\pi^2/6$, $\zeta_3 = 1.202056903 \ldots$,
$\zeta_4 = \pi^4/90$ and $\zeta_5 = 1.036927755 \ldots$).  
The relation between the $\overline{\mbox{\rm MS}}$ and {\it pole} masses
is known to three loops \cite{TA:81,pole}.

Using the $\beta$ function to trade the dependence on $\mu$ by $\alpha_s$,
the solution of the differential equation \eqn{eq:gamma} is easily found to be
$$
m_q(\mu^2) = m_q(\mu_0^2)\; \exp{
 \left\{ -\int_{\alpha_s(\mu_0^2)}^{\alpha_s(\mu^2)}
{d\alpha_s\over\alpha_s} \, {\gamma(\alpha_s)\over\beta(\alpha_s)}\right\}} 
\approx m_q(\mu_0^2)\; \left( {\alpha_s(\mu^2)\over \alpha_s(\mu_0^2)}
\right)^{-\gamma_1/\beta_1} .
$$
Since $\gamma(\alpha_s)/\beta(\alpha_s)$ is positive,
quark masses are smaller at higher energies:
\bel{eq:mZone}
m_q(1\:\mbox{\rm GeV}^2)/m_q(M_Z^2) = 2.30 \pm 0.05 \, .
\ee

Quark mass ratios are independent of renormalization conventions. This is
the reason why we have been able to fix the light quark ratios 
\eqn{eq:Weinbergratios} using chiral symmetry considerations.
Including the next-to-leading $\cO(p^4)$ chiral corrections, the ratios
of light quark masses have been determined to be \cite{LE:96}:
\bel{eq:lqm}
{m_u\over m_d} = 0.553\pm 0.043 \, ;
\qquad\qquad
{m_s\over m_d} = 18.9\pm 0.8 \, .
\ee

The quark masses have been estimated with different
methods (QCD sum rules, lattice, $\tau$ decays, \ldots). Although their
precise numerical values have been rather controversial in the past, some
consensus starts to emerge. 
%Table~\ref{tab:mq} shows my personal view on 
The present status is shown in Table~\ref{tab:mq}. 
The heavy quark masses are given at the quark--mass
scale itself, \ie\ $m_q(m_q^2)$, while for light quarks a reference
scale $\mu_0 = 1$ GeV has been chosen,
$\overline{m}_q \equiv m_q(1\:\mbox{\rm GeV}^2)$.

%%%%%%%%%%%%%%% TABLE Quarm Masses %%%%%%%%%%%%
\begin{table}[htb]
\caption{Present values of the quark masses in GeV. 
%The light quark masses are given at the scale $\mu_0 = 1$ GeV, \ie\
$\overline{m}_q \equiv m_q(1\:\mbox{\rm GeV}^2)$.}
\label{tab:mq}
\vspace{0.2cm}
\begin{center}
\begin{tabular}{|c|c|c|}
\hline
$\overline{m}_u = 0.0046 \pm 0.0009$ &
$\overline{m}_d = 0.0082 \pm 0.0016$ &
$\overline{m}_s = 0.164 \pm 0.033$
\\ \hline
$m_c(m_c^2) = 1.2 \pm 0.1$ &
$m_b(m_b^2) = 4.2 \pm 0.1$ &
$m_t(m_t^2) = 165 \pm 5$
\\ \hline
\end{tabular}
\end{center}
\end{table}
%%%%%%%%%%%%%%%%%%%%%%%%%%%%%%%%%%%%%%%%%%%%%%%%

The up and down quark masses have been fixed from the QCD sum rule
estimate \cite{BPR:95}, 
$\overline{m}_u + \overline{m}_d = 12.8\pm 2.5$ MeV, using
the $m_u/m_d$ ratio \eqn{eq:lqm}. 
The quoted strange quark mass corresponds to the recent 
$\cO(\alpha_s^3)$ determination
from the Cabbibo--suppressed $\tau$ decay width \cite{PP:99},
$m_s(m_\tau^2) = 119 \pm 24$ MeV.
It agrees with previous QCD sum rules results \cite{QCDSRms} and
recent \cite{latticems} lattice estimates (smaller results are
found by other lattice groups \cite{latticems2}).
The resulting ratio $\overline{m}_s/\overline{m}_d = 20\pm 6$ is
in nice agreement with the chiral value \eqn{eq:lqm}.

The value given for the charm quark mass  reflects the present
lattice and QCD sum rules estimates \cite{NA:99}. 
The bottom quark mass has been extracted from the behaviour of  
$\sigma(e^+e^-\to b\bar b)$ near threshold \cite{bmass}, and from a recent
unquenched lattice calculation of the $B$ meson binding energy
\cite{latticemb}.
When evolved to the $Z$ mass scale, it implies
$m_b(M_Z^2) = 2.9 \pm 0.1$, in
nice agreement with the
value $m_b(M_Z^2) = 2.61 \pm 0.55$ obtained \cite{LEPmb} from the
3-jet $Z\to b\bar b G$ production rate \cite{German}; this provides evidence
for the running of quark masses.

Finally, the top quark mass given in the Table is the
value measured at Fermilab \cite{mt}, 
$m_t = 174.3 \pm 5.1$ GeV, assuming that
it corresponds to a {\it pole mass} definition. It is in good agreement
with the result
obtained from electroweak fits at the $Z$ peak \cite{LEPEWG:99}.

\section{Summary}

Strong interactions are characterized by three basic properties:
asymptotic freedom, confinement and dynamical chiral symmetry breaking.

Owing to the gluonic self-interactions, the QCD coupling becomes
smaller at short distances, leading indeed to an asymptotically--free
quantum field theory.
Perturbation theory can then be applied at large momentum transfers.
The resulting predictions have achieved a remarkable success,
explaining a wide range of phenomena in terms of a single coupling.
The running of $\alpha_s$ has been experimentally tested at
different energy scales, confirming the predicted QCD behaviour.

The growing of the running coupling at low energies
makes very plausible 
that the QCD dynamics generates the required confinement
of quarks and gluons into colour--singlet hadronic states.
A rigorous proof of this property is, however, still lacking.
At present, the dynamical details of hadronization are completely unknown.

Non-perturbative tools, such as QCD sum rules and lattice calculations,
provide indirect evidence that QCD also implies the proper pattern
of chiral symmetry breaking.
The results obtained so far support the existence of a non-zero
$q$-$\bar q$ condensate in the QCD vacuum, which dynamically breaks
the chiral symmetry of the Lagrangian.
However, a formal understanding of this
phenomena has only been achieved in some approximate limits.

Thus, we have at present an overwhelming experimental and theoretical
evidence that the $SU(3)_C$ gauge theory correctly describes the hadronic world.
This makes QCD
the established theory of the strong interactions.
Nevertheless,
the non-perturbative nature of its low-energy limit 
is still challenging our theoretical capabilities.

\section*{Acknowledgments}
I would like to thank the organizers for the charming atmosphere of this
school, and the students for their many interesting questions and
comments.
This work has been supported in part by the ECC, TMR Network EURODAPHNE
(ERBFMX-CT98-0169), and by DGESIC (Spain) under grant No. PB97-1261.

%\newpage
%%%%%%%%%%%%%%%%%%%%%%%%%%%%%%%%%%%%%%%%%%%%%%%%%%%%%%%%%%%%%%%%%%%%

\renewcommand{\theequation}{A.\arabic{equation}}
\setcounter{equation}{0}
%\appendix
\section*{\boldmath Appendix A: $SU(N)$ Algebra}

$SU(N)$ is the group of $N\times N$ unitary matrices, 
$U U^\dagger = U^\dagger U =1$, with $\det U=1$. 
The generators of the $SU(N)$ algebra,
$T^a$ ($a=1,2,\ldots,N^2-1$), are hermitian,
traceless matrices satisfying the commutation relations
\bel{eq:T_com}
[T^a, T^b] \, = \, i f^{abc}\, T^c \, ,
\ee
the structure constants $f^{abc}$ being real and totally antisymmetric.

The fundamental representation $T^a = \lambda^a/2$ is $N$--dimensional.
For $N=2$, $\lambda^a$ are the usual Pauli matrices, while
for $N=3$, they correspond to the eight Gell-Mann matrices:
\beqn\label{eq:GM_matrices}
\lambda^1 =\left( \bath 0 & 1 & 0 \\ 1 & 0 & 0 \\ 0 & 0 & 0 \ea \right) ,
\quad\,
\lambda^2 & = &
\left( \bath 0 & -i & 0 \\ i & 0 & 0 \\ 0 & 0 & 0 \ea \right) ,
\quad
\lambda^3 = \left( \bath 1 & 0 & 0 \\ 0 & -1 & 0 \\ 0 & 0 & 0 \ea \right) ,
\quad  
\lambda^4 = \left( \bath 0 & 0 & 1 \\ 0 & 0 & 0 \\ 1 & 0 & 0 \ea \right) ,
 \no\\   && \\ 
\lambda^5 =   
\left( \bath 0 & 0 & -i \\ 0 & 0 & 0 \\ i & 0 & 0 \ea \right) \! ,
\;\;
\lambda^6 & = &
\left( \bath 0 & 0 & 0 \\ 0 & 0 & 1 \\ 0 & 1 & 0 \ea \right)\! , \;\;
\lambda^7 = \left( \bath 0 & 0 & 0 \\ 0 & 0 & -i \\ 0 & i & 0 \ea \right)
\! , \;\;
\lambda^8 = 
{1\over\sqrt{3}}
\left( \bath 1 & 0 & 0 \\ 0 & 1 & 0 \\ 0 & 0 & -2 \ea \right) \! . \no
\eeqn
They satisfy the anticommutation relation
\bel{eq:anticom}
\left\{\lambda^a,\lambda^b\right\} \, = \,
{4\over N} \,\delta^{ab} \, I_N \,
+ \, 2 d^{abc} \, \lambda^c \, ,
\ee
where $I_N$ denotes the $N$--dimensional unit matrix and the constants
$d^{abc}$ are totally symmetric in the three indices.

For $SU(3)$, the only non-zero (up to permutations) 
$f^{abc}$ and $d^{abc}$ constants are
\beqn\label{eq:constants}
&&{1\over 2} f^{123} = f^{147} = - f^{156} = f^{246} = f^{257} = f^{345}
= - f^{367} 
\CR && \phantom{{1\over 2} f^{123}}
= {1\over\sqrt{3}} f^{458} = {1\over\sqrt{3}} f^{678} = 
{1\over 2}\, , 
\\
&&d^{146} = d^{157} = -d^{247} = d^{256} = d^{344} =
d^{355} = -d^{366} = - d^{377} = {1\over 2}\, ,  
\CR
&&d^{118} = d^{228} = d^{338} = -2 d^{448} = -2 d^{558} = -2 d^{688} =
-2 d^{788} = -d^{888} = {1\over \sqrt{3}}\, . 
\no
\eeqn

The adjoint representation of the $SU(N)$ group is given by
the $(N^2-1)\!\times\! (N^2-1)$ matrices  
$(T^a_A)_{bc} \equiv - i f^{abc}$.
The relations
\beqn\label{eq:invariants}
{\rm Tr}\left(\lambda^a\lambda^b\right) =  4 \, T_F \, \delta_{ab}
\, , \qquad\qquad\quad\quad && T_F = {1\over 2} \, ,
\no\\
\left(\lambda^a\lambda^a\right)_{\alpha\beta} = 4
C_F \, \delta_{\alpha\beta} \, , \qquad\qquad\quad\quad 
&& C_F = {N^2-1\over 2N} \, ,
\\ \;
{\rm Tr}(T^a_A T^b_A) = f^{acd} f^{bcd} = C_A \,\delta_{ab} \, , 
\qquad\qquad
&& C_A = N \, , \qquad\no
\eeqn
define the $SU(N)$ invariants $T_F$, $C_F$ and $C_A$.
Other useful properties are:
$$
\left(\lambda^a\right)_{\alpha\beta} 
\left(\lambda^a\right)_{\gamma\delta} 
= 2 \delta_{\alpha\delta}\delta_{\beta\gamma}
 -{2\over N} \delta_{\alpha\beta}\delta_{\gamma\delta} \, ; 
\qquad\qquad
{\rm Tr}\left(\lambda^a\lambda^b\lambda^c\right)
 =  2 (d^{abc} + i f^{abc})\, ;
$$
$$
{\rm Tr}(T^a_A T^b_A T^c_A)  =  i \, {N\over 2} f^{abc} \, ;
\qquad \sum_b d^{abb} = 0\, ; \qquad
d^{abc} d^{ebc}  =  \left( N - {4\over N}\right) \delta_{ae} \, ; 
$$
$$
f^{abe} f^{cde} + f^{ace} f^{dbe} + f^{ade} f^{bce} = 0 \, ; 
\qquad\quad\;
f^{abe} d^{cde} + f^{ace} d^{dbe} + f^{ade} d^{bce} = 0 \, . 
$$
%

%%%%%%%%%%%%%%%%%%%%%%%%%%%%%%%%%%%%%%%%%%%%%%%%%%%%%%%%%%%%%%%%%%%%

\renewcommand{\theequation}{B.\arabic{equation}}
\setcounter{equation}{0}
\section*{Appendix B: Gauge Fixing and Ghost Fields}

The fields $G^\mu_a$ have 4 Lorentz degrees of freedom, 
while a massless spin-1 gluon has 2 physical polarizations only.
Although gauge invariance makes the additional degrees of freedom irrelevant,
they give rise to some technical complications when quantizing the gauge
fields. 

The canonical momentum associated with $G^\mu_a$,
$\Pi^a_\mu(x)\equiv
\delta\cL_{\mbox{\rms QCD}}/\delta (\partial_0G^\mu_a)
= G^a_{\mu 0}$, 
vanishes identically for $\mu=0$.
The standard commutation relation
\bel{eq:quantization}
\left[ G^\mu_a(x), \Pi^\nu_b(y) \right] \delta(x^0-y^0) \, = \,
i g^{\mu\nu} \delta^{(4)}(x-y) \, ,
\ee
is then meaningless for $\mu=\nu=0$. In fact, the field $G^0_a$ is just a
classical quantity, since it commutes with all the other fields.
This is not surprising, since we know that there are 2 unphysical components
of the gluon field, which should not be quantized. Thus, we could just impose
two gauge conditions, such as $G^0_a = 0$ and $\vec\nabla \vec G_a = 0$,
to eliminate the 2 redundant degrees of freedom,
and proceed working with the physical gluon polarizations only. However,
this is a (Lorentz) non-covariant procedure, which leads to a very
awkward formalism. Instead, one can impose a Lorentz--invariant gauge
condition, such as $\partial_\mu G^\mu_a = 0$. The simplest way to
implement this is to add to the Lagrangian the gauge-fixing term
\bel{eq:L_GF}
\cL_{\mbox{\rms GF}} \, = \, -{1\over 2\xi} \, (\partial^\mu G_\mu^a)\,
 (\partial_\nu G^\nu_a) \, ,
\ee
where $\xi$ is the so-called gauge parameter.
The 4 Lorentz components of the canonical momentum
\bel{eq:canonical}
\Pi^a_\mu(x) \equiv {\delta\cL_{\mbox{\rms QCD}}\over\delta 
(\partial_0 G^\mu_a)}
= G^a_{\mu 0} - {1\over \xi}\, g_{\mu 0}\, (\partial^\nu G_\nu^a)
\ee
are then non-zero, and one can develop a covariant quantization formalism.
Since \eqn{eq:L_GF} is a quadratic $G_a^\mu$ term, it modifies the gluon
propagator:
$$
\langle 0 | T[G^\mu_a(x) G^\nu_b(y)]|0\rangle =
i \delta_{ab} \int {d^4k\over (2\pi)^4} {\e^{-ik(x-y)}\over k^2 + i\varepsilon}
\left\{ -g^{\mu\nu} + (1-\xi) {k^\mu k^\nu\over k^2 + i\varepsilon}\right\}
\, .
$$
Notice, that the propagator is not defined for $\xi=\infty$, \ie\ in
the absence of the gauge-fixing term \eqn{eq:L_GF}.

In QED, this gauge-fixing procedure is enough for making a consistent 
quantization of the theory. The initial gauge symmetry of the Lagrangian
guarantees that the redundant photon polarizations do not generate
non-physical contributions to the scattering amplitudes, and the final
results are independent of the arbitrary gauge parameter $\xi$.
In non-abelian gauge theories, like QCD, a second problem still remains.

%%%%%%%%%%%%%%%
\begin{figure}[htb] 
\myfigure{\epsfysize =3cm \epsfbox{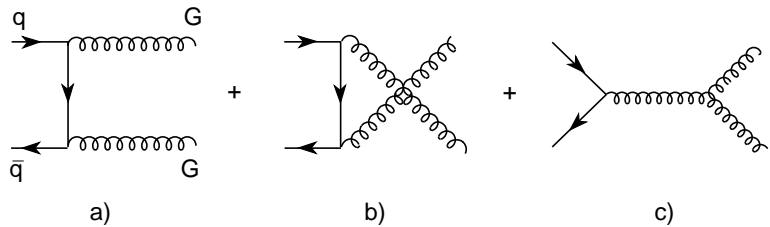}}
\vspace{-0.2cm}
\caption{Tree-level Feynman diagrams contributing to $q\bar q\to GG$.}
\label{fig:qqGG}
\end{figure}
%%%%%%%%%%%%%%%

Let us consider the scattering process $q\bar q\to G G$, which proceeds
through the three Feynman graphs shown in Fig.~\ref{fig:qqGG}. The scattering
amplitude has the general form
$T = J^{\mu\mu'} \varepsilon_{\mu\phantom{'}}^{(\lambda)} 
\varepsilon_{\mu'}^{(\lambda')}$.
The probability associated with the scattering process, 
\bel{eq:prob}
\cP \sim {1\over 2} J^{\mu\mu'} (J^{\nu\nu'})^\dagger \,
\sum_\lambda \varepsilon_\mu^{(\lambda)}\varepsilon_\nu^{(\lambda)*}\,
\sum_{\lambda'} \varepsilon_{\mu'}^{(\lambda')}\varepsilon_{\nu'}^{(\lambda')*}
\, ,
\ee
involves a sum over the final gluon polarizations.
One can easily check that the physical probability $\cP_T$, where only the
two transverse gluon polarizations are considered in the sum, is different
from the covariant quantity $\cP_C$, which includes a sum over all
polarization components:  $\cP_C > \cP_T$.
In principle, this is not a problem because only $\cP_T$ has physical meaning;
we should just sum over the physical transverse polarizations to get the
right answer. However, the problem comes back at higher orders.

%%%%%%%%%%%%%%%
\begin{figure}[hbt]
\myfigure{\epsfysize =3cm \epsfbox{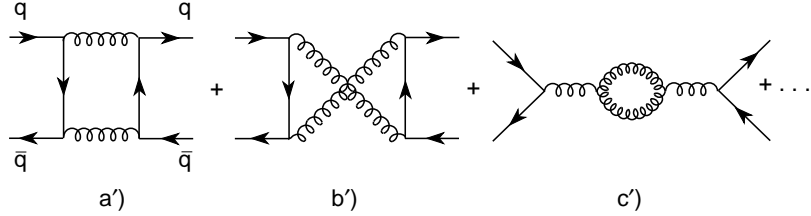}}
\vspace{-0.2cm}
\caption{1-loop diagrams contributing to $q\bar q\to q\bar q$.}
\label{fig:loop}
\end{figure}
%%%%%%%%%%%%%%%

The covariant gluon propagator contains the 4 polarization
components; therefore higher-order graphs such as the ones in 
Fig.~\ref{fig:loop} get unphysical contributions from the longitudinal and
scalar gluon polarizations propagating along the internal gluon lines.
%%%%%%%%%%%%\footnote{
The absorptive part of these 1-loop graphs (\ie\ the imaginary part obtained 
putting on-shell the two gluon lines within the loop) 
is equal  to $|T(q\bar q\to GG)|^2$.
Thus, these loops suffer the same probability problem than the tree-level
$q\bar q\to GG$ amplitude.
%%%%%%%%%%%%}.
The propagation of unphysical gluon components implies then a violation
of unitarity (the two fake polarizations contribute a positive probability).

In QED this problem does not appear because the gauge-fixing condition
$\partial^\mu A_\mu=0$ still leaves a residual gauge invariance
under transformations satisfying $\Box\theta = 0$. This guarantees that
(even after adding the gauge-fixing term) the electromagnetic current 
is conserved, \ie\ 
$\partial_\mu J^\mu_{\mbox{\rms em}}=\partial_\mu (eQ\bar\Psi\gamma^\mu\Psi) = 
0$.
If one considers the $e^+e^-\to\gamma\gamma$ process, which proceeds through
diagrams identical to a) and b) in Fig.~\ref{fig:qqGG}, 
current conservation implies
$k_\mu J^{\mu\mu'} = k'_{\mu'} J^{\mu\mu'} = 0$, where $k_\mu$ and $k'_{\mu'}$
are the momenta of the photons with polarizations $\lambda$ and $\lambda'$,
respectively (remember that the interacting vertices contained in 
$J^{\mu\mu'}$ are in fact the corresponding electromagnetic currents).
As a consequence, the contributions from the
scalar and longitudinal photon  polarizations vanish
and, therefore, $\cP_C = \cP_T$.

The reason why $\cP_C\not=\cP_T$ in QCD stems from the third diagram in
Fig.~\ref{fig:qqGG}, involving a gluon self-interaction.
Owing to the non-abelian character of the $SU(3)_C$ group, the gauge-fixing
condition $\partial_\mu G^\mu_a=0$ does not leave any residual 
invariance\footnote{
%%%%%%%%
To maintain $\partial_\mu (G^\mu_a)'=0$ after the gauge transformation
\eqn{eq:inf_transf}, one would need
$\Box \delta\theta_a = g_s f^{abc}\partial_\mu(\delta\theta_b) G^\mu_c$,
which is not possible because $G^\mu_c$ is a quantum field.}.
%%%%%%%%% 
Thus, $k_\mu J^{\mu\mu'}\not= 0$.
% since a conserved current does not exist.

%%%%%%%%%%%%%%%
\begin{figure}[hbt]
\myfigure{\epsfysize =3cm \epsfbox{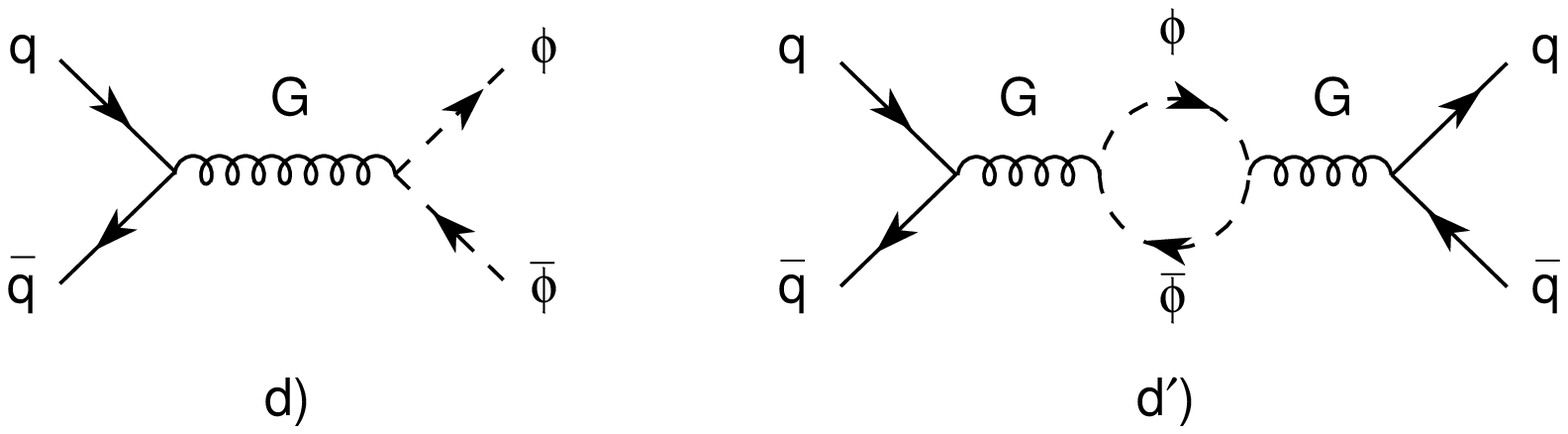}}
\caption{Feynman diagrams involving the ghosts.}
\label{fig:ghosts}
\end{figure}
%%%%%%%%%%%%%%%

Again, the problem could be solved adopting a non-covariant quantization where
only the physical transverse polarizations propagate; but the resulting
formalism would be awful and very inconvenient for performing 
practical calculations.
A more clever solution    %,DB:64}
consist \cite{FE:63} in adding additional unphysical fields,
the so-called {\it ghosts}, with a coupling to the gluons
such that exactly cancels {\it all} unphysical contributions from 
the scalar and longitudinal gluon polarizations.
Since a positive fake probability has to be cancelled, one needs fields
obeying the wrong statistics (\ie\ of negative norm)
and thus giving negative probabilities.
The magic cancellation is achieved by adding to the Lagrangian the
Faddeev--Popov term \cite{FP:67},
\bel{eq:L_FP}
\cL_{\mbox{\rms FP}} \, = \, -\partial_\mu\bar\phi_a D^\mu\phi^a \, , \qquad
\quad D^\mu\phi^a\equiv \partial^\mu\phi^a - g_s f^{abc} \phi^b G^\mu_c \, ,
\ee
where $\bar\phi^a$, $\phi^a$ ($a=1,\ldots,N_C^2-1$) is a set of 
anticommuting (\ie\ obeying the Fermi--Dirac statistics), massless,
hermitian, scalar fields. 
The covariant derivative $D^\mu\phi^a$ contains the needed coupling to the
gluon field. One can  easily check that
diagrams d) and d') in Fig.~\ref{fig:ghosts} exactly cancel the unphysical
contributions from diagrams c) and c') of Figs.~\ref{fig:qqGG} and
\ref{fig:loop}, respectively; so that finally $\cP_C=\cP_T$.
Notice, that the Lagrangian \eqn{eq:L_FP} is necessarily not hermitian,
because one needs to introduce an explicit violation of unitarity 
to cancel the unphysical probabilities and restore the unitarity of the
final scattering amplitudes.

The exact mechanism giving rise to the $\cL_{\mbox{\rms FP}}$ term
can be understood in a simpler way using the more powerful
path-integral formalism, which is beyond the scope of these lectures.
The only point I would like to emphasize here, is that the addition of
the gauge-fixing and Faddeev--Popov Lagrangians is just a mathematical
trick, which allows to develop a simple covariant formalism, and
therefore a set of simple Feynman rules, making easier to perform
explicit calculations.

\end{document}